\documentclass[a4paper,12pt]{article}
\date{}
\usepackage{graphicx}
\usepackage{amsmath}
\usepackage{amstext}
\usepackage{amsbsy}
\usepackage{txfonts}
\usepackage{caption}
\usepackage{subcaption}
\usepackage{overpic}
\usepackage{booktabs}
\usepackage{float}
\usepackage{authblk}
\usepackage{multirow}
\usepackage{hyperref}
\usepackage[labelfont=bf]{caption} 
\usepackage{cite}
\usepackage[a4paper, left=0.9in, right=0.9in, top=1in, bottom=1in]{geometry}
\usepackage{comment}
\usepackage{array}
\usepackage{color,url}
\newcolumntype{C}[1]{>{\centering\arraybackslash}p{#1}}

\renewcommand{\figurename}{Fig.}

\title{High-Fidelity Data-Driven Dynamics Model for Reinforcement Learning-based Control in HL-3 Tokamak}
\author{Niannian Wu$^{1,2,*}$, Zongyu Yang$^{2,*}$, Rongpeng Li$^{1,*,+}$, Ning Wei$^{3,*}$, Yihang Chen$^{2}$, Qianyun Dong$^{1,2}$, Jiyuan Li$^{2}$, Guohui Zheng$^{2}$, Xinwen Gong$^{2}$, Feng Gao$^{3}$, Bo Li$^{2}$, Min Xu$^{2}$, Zhifeng Zhao$^{3,1,+}$, Wulyu Zhong$^{2,+}$}
\affil{1. Zhejiang University, Hangzhou 310058, China}
\affil{2. Southwestern Institute of Physics, Chengdu 610043, China}
\affil{3. Zhejiang Lab, Hangzhou 311500, China}
\affil{$+$ Corresponding author (s). Email(s): lirongpeng@zju.edu.cn,  zhaozf@zhejianglab.com, zhongwl@swip.ac.cn.}
\affil{$*$ these authors contributed equally to this work.}

\begin{document}
\renewcommand\footnotemark{}
\renewcommand\footnoterule{}
\maketitle
\begin{abstract}
The success of reinforcement learning (RL)-based control in tokamaks, an emerging technique for controlled nuclear fusion with improved flexibility, typically requires substantial interaction with a simulator capable of accurately evolving the high-dimensional plasma state.
Compared to first-principle-based simulators, whose intense computations lead to sluggish RL training, we devise an effective method to acquire a fully data-driven simulator, by mitigating the arising compounding error issue due to the underlying autoregressive nature. With high accuracy and appealing extrapolation capability, this high-fidelity dynamics model subsequently enables the rapid training of a qualified RL agent to directly generate engineering-reasonable actuator commands, aiming at the desired long-term targets of plasma configuration. Together with a surrogate model for Equilibrium Fitting code based on neural network, named EFITNN, the RL agent successfully maintains a $400$-ms, $1$ kHz trajectory control with accurate waveform tracking of plasma current and last closed flux surface on the HL-3 tokamak. Furthermore, it also demonstrates the feasibility of zero-shot adaptation to changed triangularity targets, confirming the robustness of the developed data-driven dynamics model. Our work underscores the advantage of fully data-driven dynamics models in yielding RL-based trajectory control policies at a sufficiently fast pace, an anticipated engineering requirement in daily discharge practices for the upcoming ITER device.

\vspace{0.1cm}
\end{abstract}

\section{Introduction}

The objective of the tokamak, a torus-shaped nuclear fusion device, is to provide sustainable energy by effectively confining high-temperature plasmas using magnetic fields. The stable confinement of plasma in tokamak depends significantly on well-calibrated control approaches, which can be developed through a thorough understanding of the underlying dynamics \cite{he2015coherent, fortov2005complex, yuan2013plasma}. Rather than providing an exhaustive interpretation of fundamental physics, control-oriented dynamics models deliver significant advantages in simplicity with reasonable accuracy \cite{walker2015iter}, assisting the precise manipulation of various actuators toward desired configurations. For instance, 
the reliable prediction of rapidly growing plasma instabilities contributes to proactively suppressing disruptions \cite{seo_avoiding_2024,maljaars2015control,felici2018real,van2023scenario}, while an accurate long-term evolution of the plasma dynamics supports Reinforcement Learning (RL)-based controller to meticulously regulate high-dimensional, mutually coupled magnetic coils towards target plasma current and shape \cite{degrave2022magnetic,TRACEY2024114161}. 

Classically, a high-fidelity, autoregression-capable dynamics model \cite{seo2021feedforward,seo_development_2022} can be built on first-principles-based computations. A notable example is the Forward Grad–Shafranov Evolutive (FGE) model, which solves the free-boundary equilibrium while accounting for the plasma current evolution coupled to controller dynamics \cite{carpanese_development_2021}. Through accurately synthesizing high-dimensional magnetic flux and local magnetic field measurements as outputs of the tokamak sensors, this model enables the offline training of RL-based magnetic controller \cite{degrave2022magnetic}. Although this physics-driven approach can give highly detailed simulations, an over-reliance on such simulators without incorporating sufficient real-world data may reduce the generalizability of RL policies \cite{zhang2024real}. Additionally, each training step has to await the FGE simulator for the cumbersome computations of lumped-parameter equations \cite{degrave2022magnetic}, making the training of RL policies possibly take hours \cite{TRACEY2024114161}. In contrast, ITER, the International Thermonuclear Experimental Reactor, is expected to be capable of executing the next pulse within a minimum of $30$ minutes  \cite{iter_research-plan2024}. Despite progress to accelerate training, the physics-driven simulator is still not competent for training RL-based control policies at a sufficiently fast pace \cite{TRACEY2024114161}, an anticipated engineering requirement in daily discharge practices for ITER. 

Compared to first-principles-based approaches, historical discharge data offers the potential to develop models with significantly faster speed. For example, experimental data facilitates the adoption of linearized system identification techniques to approximately reproduce the response of coupled plasma profiles under variations of given actuators \cite{moreau2013integrated,zhang2024real}. Nevertheless, the validity of the linearized dynamics model is limited to the vicinity of the plasma equilibrium that needs to be tracked \cite{Moreau_plasma_2011}. An alternative approach resorts to Artificial Intelligence (AI) models, as Deep Neural Network (DNN)-based models demonstrate satisfactory accuracy \cite{kates-harbeck_predicting_2019,dong_deep_2021,li_simulation_2022,wan_predict_2024,zhu_datadriven_2022} with preliminary dynamics modeling results \cite{seo_development_2022,seo2021feedforward,seo_avoiding_2024}. Prominently, a DNN model that competently forecasts the future tearing likelihood from various plasma profiles and tokamak actuations, has successfully contributed to the prediction-driven planning of a high-level policy to yield desired beam power and plasma triangularity target for tearing avoidance \cite{seo_avoiding_2024}.  
Though such a process disentangles the predictive power of the dynamics model from the policy learning procedure, 
building a high-fidelity, data-driven simulator for training a high-dimensional, high-frequency RL controller remains challenging. Unless optimizations are adopted to make
long-range predictions, the autoregressive nature of such simulators can lead to compounding errors, which gradually accumulate over long-term multiple-step evolution, resulting in significant discrepancies and potentially catastrophic outcomes \cite{asadi2018lipschitz}.

\begin{figure}[!p]
    \centering
    \includegraphics[width=0.97\textwidth]{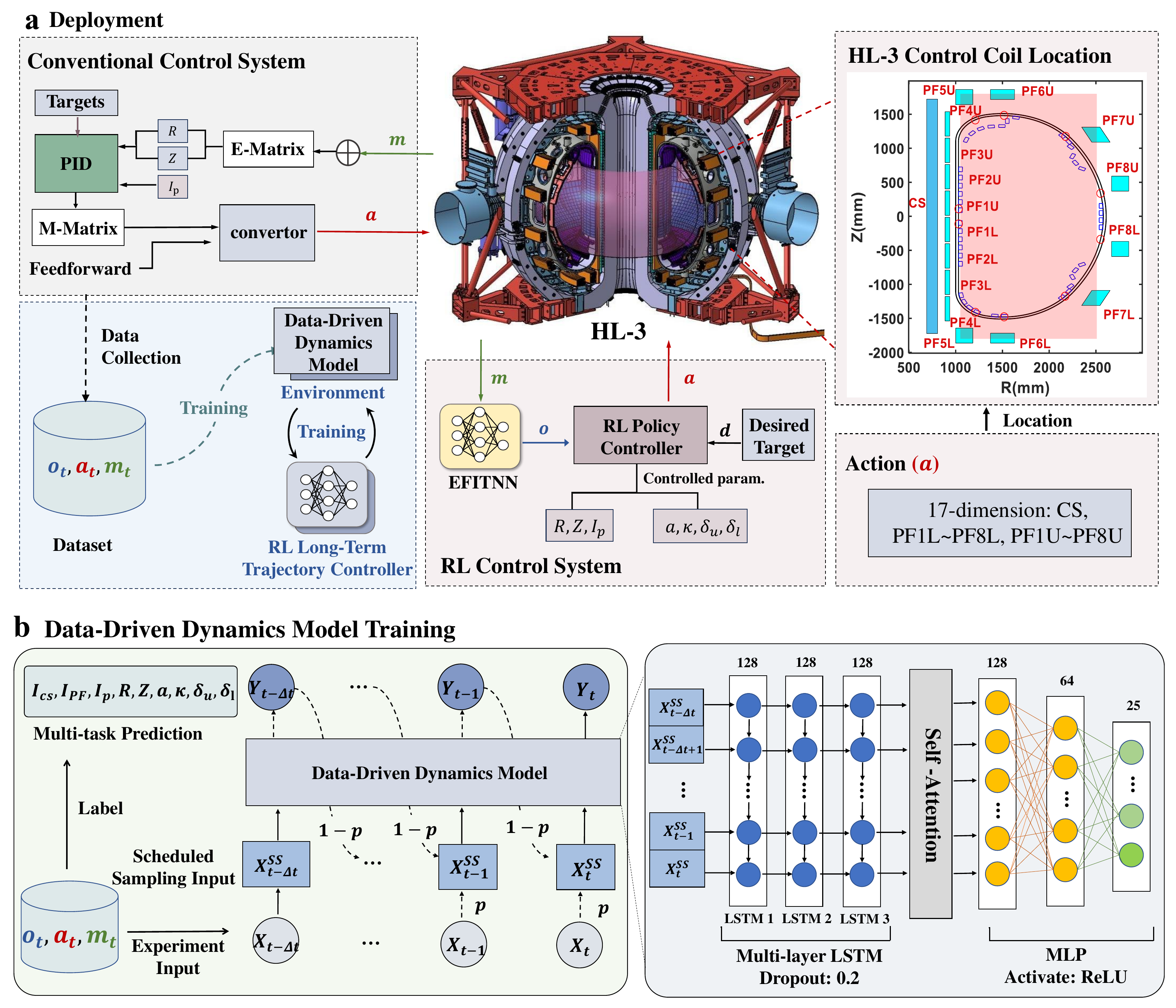}
    \caption{\textbf{The overall architecture of the plasma current and shape control in the HL-3 tokamak}. \textbf{a}, Deployment schematic for plasma control on the HL-3 tokamak, illustrating two control paradigms. The Conventional Control System utilizes a Proportional-Integral-Derivative (PID) controller. It operates on the error between desired targets and a measured plasma state derived from the direct diagnostic measurements $\mathbf{m}$ via the M-Matrix, which calculates the $R$ and $Z$ positions for control, to ultimately generate commands that contribute to the final actuator voltages $\mathbf{a}$. The Reinforcement Learning (RL) Control System employs a trained policy controller. This controller receives its state $\mathbf{o}$ from the diagnostic measurements $\mathbf{m}$ processed by the EFITNN model. This state $\mathbf{o}$ is composed of the plasma current $I_\mathrm{p}$ as well as six shape and position parameters: the minor radius $a$, elongation $\kappa$, upper triangularity $\delta_\mathrm{u}$, lower triangularity $\delta_\mathrm{l}$, and the radial and vertical positions, $R$ and $Z$, of the plasma geometric center. The controller then takes this state $\mathbf{o}$ and the desired target state $\mathbf{d}$ as inputs to directly generate the action $\mathbf{a}$, which represents the voltage commands for the $17$ magnetic coils. Historical data from the conventional system, including direct diagnostic measurements $\mathbf{m_t}$, derived state observations $\mathbf{o_t}$, and the executed actuator commands $\mathbf{a_t}$, are collected to train the data-driven dynamics model. This model then serves as a high-fidelity simulated environment for the offline training of the RL controller. 
\textbf{b}, The proposed training diagram of the dynamics model, including scheduling sampling, multi-task adaptive weight adjustment, self-attention, and model ensembling.}
    \label{fig:diagram}
\end{figure}

To combat the long-term divergence issue, one intuitive means is to unroll the single-step prediction to incorporate multi-prediction losses during training \cite{moerland_modelbased_2023,ke_modeling_2019,hafner_learning_2019,chiappa_recurrent_2017,abbeel_learning_2004}; while another line of work lies in implicitly \cite{lambert2021learning,lambert2022investigating,mishra_prediction_2017} or explicitly \cite{asadi2018towards, char2024full} learning a specific dynamics model for every $n$-step prediction. Albeit the practicability in low-dimensional scenarios \cite{ke_modeling_2019,hafner_learning_2019,chiappa_recurrent_2017,abbeel_learning_2004,lambert2021learning,lambert2022investigating,mishra_prediction_2017, asadi2018towards}, both strategies imply significant training difficulties with limited scalability for a magnetic control that usually lasts for hundreds or even thousands of steps. Therefore, on top of single-step prediction, several techniques like ensembling \cite{abbate2021data,abbate_general_2023}, principle component analysis \cite{char2023offline} and noise layer \cite{seo_development_2022,seo2021feedforward} have been leveraged to ameliorate this issue. However, to our best knowledge, the data-driven dynamics model is still far from perfect for long-term, high-dimensional trajectory control. Inevitably, Proportional-Integral-Derivative (PID) controllers \cite{seo_development_2022,seo2021feedforward,seo_avoiding_2024}, historically logged multi-variate commands \cite{char2023offline, char2024pid}, or online Model Predictive Control (MPC) \cite{abbate_general_2023} indispensably complements for achieving localized short-term, low-level magnetic control. Nevertheless,
PID controllers imply tremendous engineering and design efforts before deployment \cite{degrave2022magnetic}, while online MPC suffers from the cumbersome online autoregression of the dynamics model for trajectory sampling \cite{maljaars2015control,abbate_general_2023}. 
Consequently, due to the rare readiness of either low-level controllers or physical simulators, the limited accuracy with only short-term guarantees in existing data-driven dynamic models hinders the learning of high-dimensional, long-term trajectory control policies.

In this paper, we aim to provide a data-driven dynamics model, that exclusively learns from the discharge logs collected through the interaction between PID controllers and the tokamak device. The dynamics model is designed to efficiently support the fast training and practical deployment of a long-term, low-level RL-based agent, capable of directly generating actuator commands. The system architecture is shown in Fig. \ref{fig:diagram}. As shown in Fig. \ref{fig:diagram}a, EFITNN \cite{zheng2024real}, a surrogate model of magnetic equilibrium reconstruction code EFIT \cite{lao1990equilibrium, lao1985reconstruction}, is leveraged to meaningfully transform the high-dimensional magnetic measurements to lower-dimensional plasma shape information while maintaining the physical interpretability and supporting real-time deployment. Subsequently, a Long Short-Term Memory Network (LSTM)-based \cite{greff_lstm_2017} dynamics model is developed to track the evolution of plasma shape and current, based on accessible zero-dimensional plasma parameters and executed actuations. Benefiting from effectively leveraging several useful techniques including scheduling sampling  \cite{bengio_scheduled_2015,venkatraman_improving_2015}, adaptive weight adjustment \cite{liebel2018auxiliary,kendall2018multi}, self-attention \cite{vaswani2017attention} and model ensembling \cite{rozante2014multi}, the trained dynamics model demonstrates exceptional generalization capability on unseen plasma current regions during training and sensitively captures shape variations in typical scenarios, such as transitioning from limiter to (advanced) divertor configuration \cite{ryutov2012snowflake}, and soft ramp-down \cite{de2017multi} during discharges. Unlike localized short-term planning \cite{seo_avoiding_2024,seo_development_2022,seo2021feedforward} shown in Supplementary Figure 1a, this data-driven dynamics model enables fully RL-based $400$-ms, $1$-kHz low-level trajectory control of up to $17$ magnetic coils, achieving the desired Last Closed Flux Surface (LCFS) \cite{han2022tracking} via six shape and position parameters; while as in Supplementary Figure 1b, the training of $250,000$ iterations completes within $22$ minutes on the customer-grade Nvidia GeForce RTX 4090 GPU, fully satisfying the anticipation for RL-based magnetic control in ITER. It even achieves zero-shot triangularity adjustment, validating the effectiveness and reliability of the data-driven dynamics model.

\section{Results}
\subsection{Construction of Data-Driven Dynamics Model}
\begin{figure}[!tb]
    \centering
    \includegraphics[width=.475\textwidth]{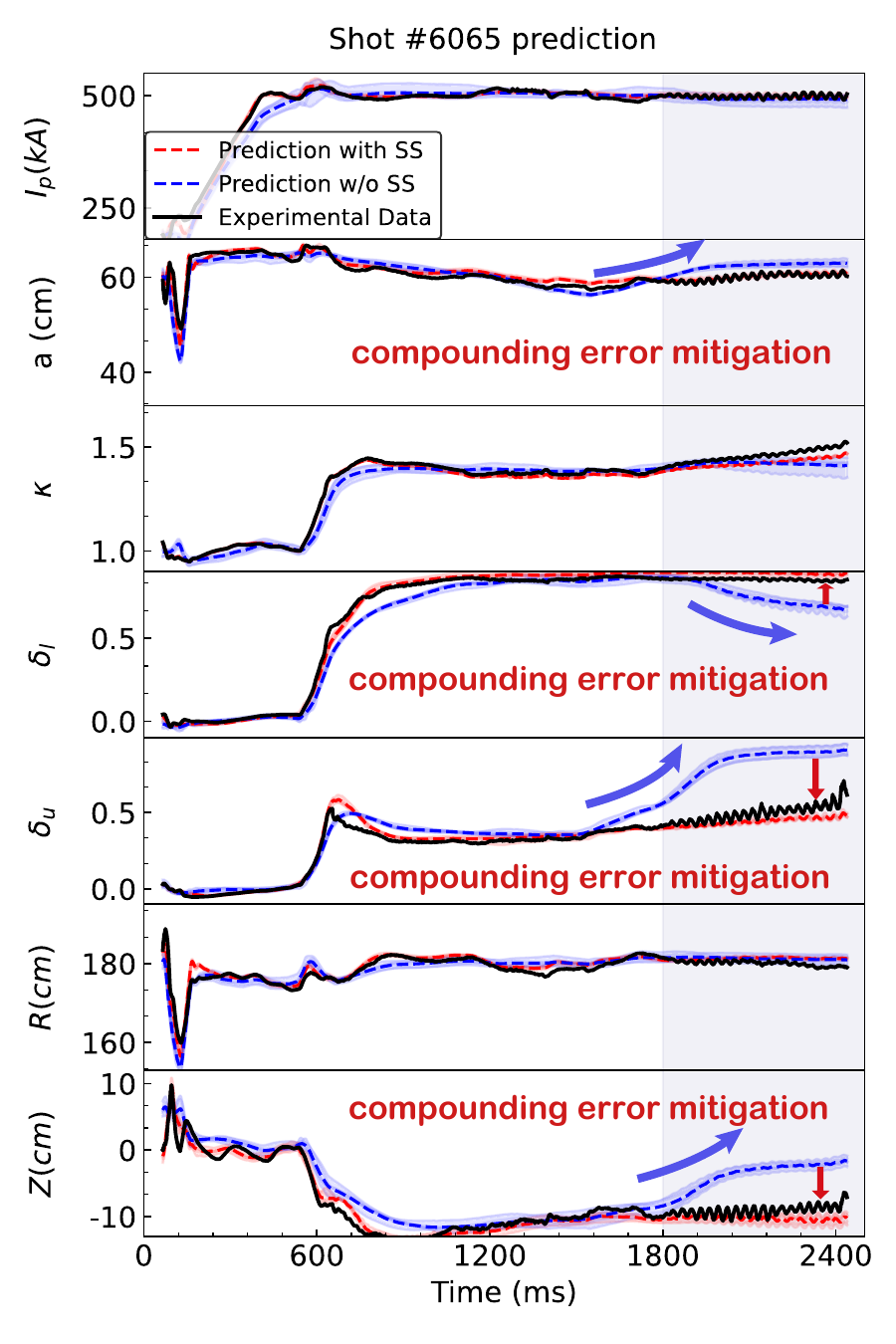}
    \caption{\textbf{The demonstration case to predict plasma shape and position parameters for shot \#6065 with and without the involvement of Scheduled Sampling.} Compared with actual discharge data (black solid line), a model trained with scheduled sampling (red dashed line) more effectively mitigates the compounding error than that without scheduled sampling (blue dashed line). The shaded area represents the results of ensembling $5$ independent models.}
    \label{fig:6065 500kA}
\end{figure}

To train a $1$-kHz RL-based control policy, it is essential to build a high-fidelity data-driven dynamics model, which can autoregressively produce accurate predictions for up to $25$ variates.
These predictive variates encompass the plasma current, $6$ plasma shape and position parameters, and $18$ channels of feedforward coil currents, which together correspond to a $44$-dimensional indispensable input (Supplementary Table 2) and support the training of an engineering-reasonable policy for controlling $17$ magnetic coil voltages. In comparison to yielding $100$ ms-averaged evolution \cite{seo_development_2022},  forecasting single-variate tearability \cite{seo_avoiding_2024}, or only tuning the total power and torque injected from the neutral beams  \cite{char2023offline}, the significantly higher resolution and a larger number of coupling actuators implies to pose considerable challenges and requires a more comprehensive design. 

\begin{figure}[!t]
    \centering
    \includegraphics[width=1.0\textwidth]{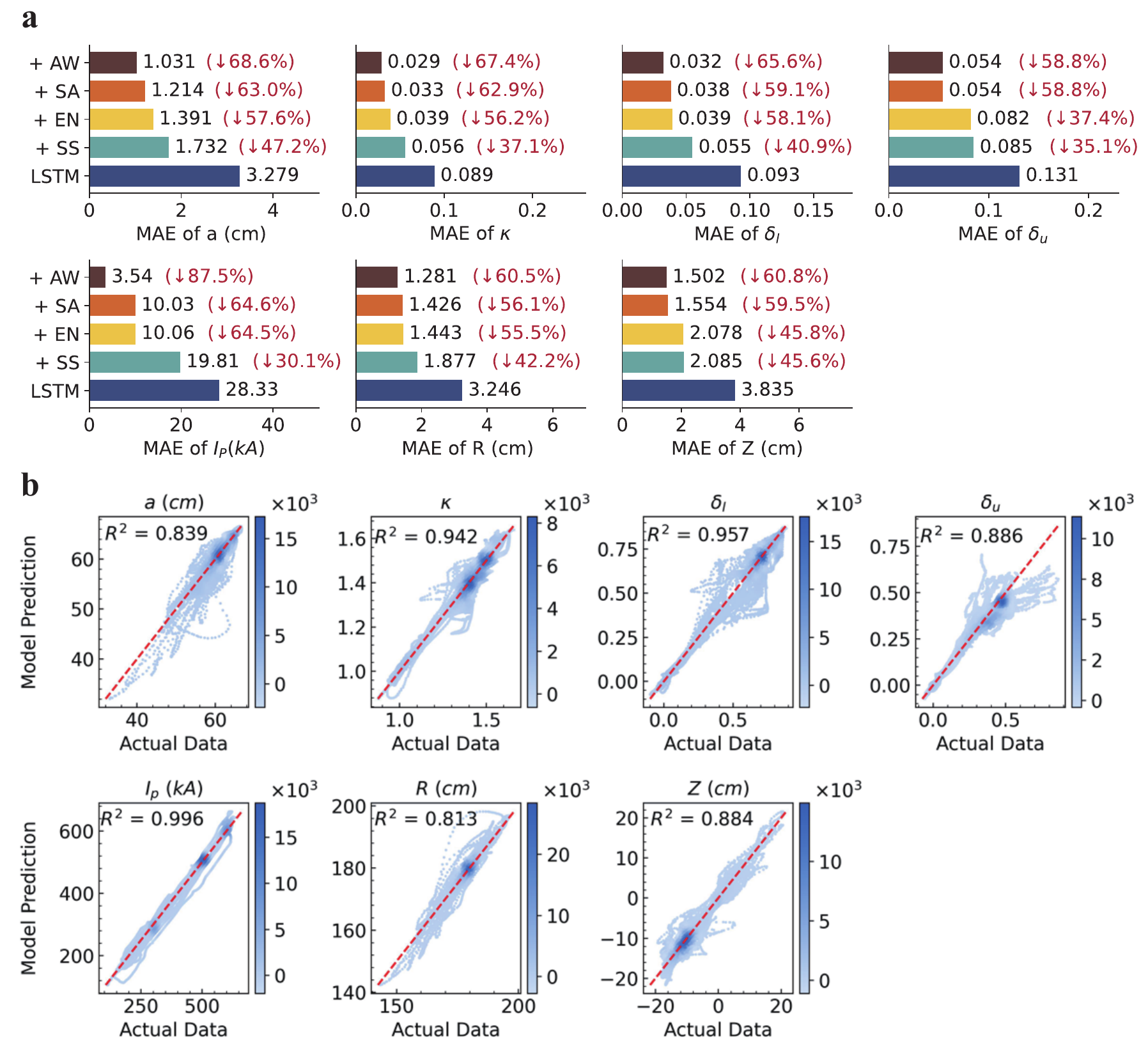}
    \caption{\textbf{Accuracy of the dynamics model.} \textbf{a}, Ablations for key techniques in terms of MAE. \textbf{SS}: Scheduled Sampling, \textbf{EN}: Ensemble, \textbf{SA}: Self-Attention, \textbf{AW}: Adaptive Weight. Each group depicted here is derived by incorporating the corresponding module onto the basis of the preceding group. For instance, the notation ``+ EN" signifies the integration of the Ensemble into the LSTM model that already incorporates scheduled sampling (``+ SS''). \textbf{b}, Goodness-of-fit of the dynamics model in terms of the coefficient of determination $R^2$. The color map indicates the density of data points and the red dashed line represents the line of perfect agreement.}
\label{fig:trick_all}
\end{figure}

During training, to mitigate the compounding errors, scheduled sampling, which blends the historical logged data and model output \cite{bengio_scheduled_2015,venkatraman_improving_2015}, is leveraged. Fig. \ref{fig:6065 500kA} shows this easy-to-implement technique ensures the training scalability and brings appealing results; in Methods, a more thorough comparison with alternative solutions such as noise layer \cite{seo_development_2022} shows the flexibility of this training approach. On this top, training the dynamics model incorporates several useful techniques. In particular, adaptive weight adjustment \cite{liebel2018auxiliary,kendall2018multi}, which simplifies the tuning of weights for combining multi-task loss functions \cite{kendall2018multi} and guarantees resilient and steady model learning \cite{liebel2018auxiliary}, is adopted to capture the implicit relationship among plasma shape and position parameters. Meanwhile, auxiliary techniques, such as self-attention \cite{vaswani2017attention} and model ensembling \cite{rozante2014multi}, further enhance accuracy through a concerted effort.
Fig. \ref{fig:trick_all}a presents the ablation study in terms of Mean Absolute Error (MAE) and demonstrates the incremental contribution of each technique to train the dynamics model. Besides, the model maintains high consistency across different tasks, achieving at least a $58.8\%$ reduction in MAE than the standard LSTM-based training. In addition, Fig. \ref{fig:trick_all}b provides the statistical measure in terms of the coefficient of determination $R^2$ and shows the autoregressive output of the dynamics model can fit the actual data well.

\subsection{Prediction Accuracy under Abrupt Shape variations}

\begin{figure}[!ht]
\centering
    \includegraphics[width=1.0\textwidth]{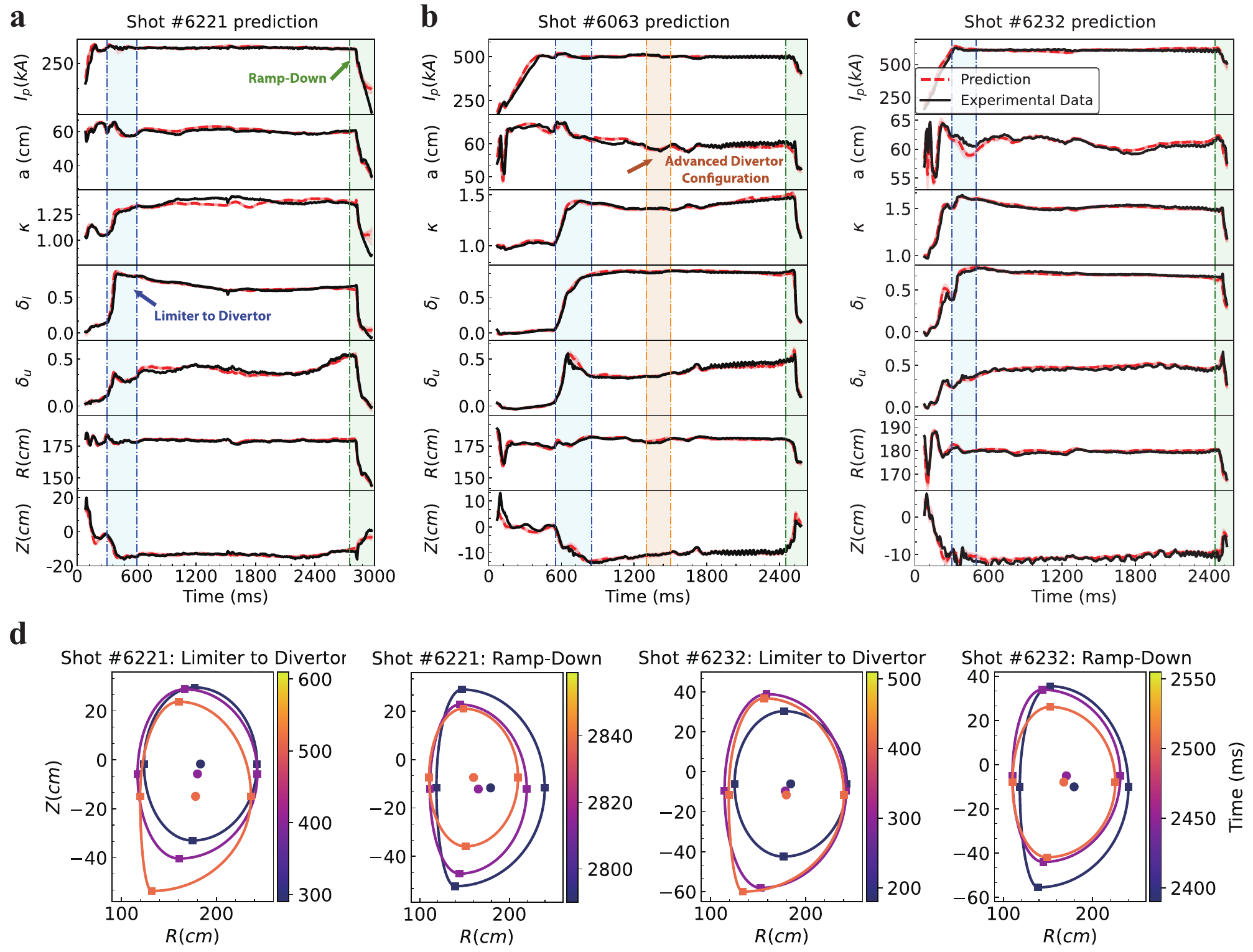}
    \caption{\textbf{The accuracy of dynamics model to predict abrupt plasma shape variations.} In panels \textbf{a}-\textbf{c}, the model's prediction (red dashed line) is compared with experimental data (black solid line). The shaded regions highlight key operational phases: the light blue region marks the transition from the limiter to divertor configuration, the light orange region indicates the advanced divertor configuration transformation phase and the light green region denotes the plasma current ramp-down phase.  \textbf{a}, Shot \#6221, an H-mode discharge with ELMs and a plasma current of $300$ kA. \textbf{b}, Shot \#6063, an advanced divertor configuration discharge with a plasma current of $500$ kA, with advanced divertor configuration undergoing a complete transformation between $1300$ ms and $1500$ ms. \textbf{c}, Shot \#6232, an ITB discharge with a plasma current of $600$ kA. \textbf{d}, The Last Closed Flux Surface (LCFS) variations correspond to the transition from limiter to divertor shape and soft ramp-down phases for shot \#6221 ($300$ kA) and shot \#6232 ($600$ kA), respectively.}
\label{fig:dynamics model results}
\end{figure}
In the previous section, we demonstrated the capability of the data-driven dynamics model in learning the high-dimensional evolution. In this section, we switch to investigate the prediction accuracy under abrupt plasma shape variations, which are commonly encountered due to the transition from limiter to divertor shape, the advanced divertor configuration transformation, or the initialization of soft ramp-down. Taking the example of shot \#6221, an H-mode discharge with edge localized modes (ELMs), Fig. \ref{fig:dynamics model results}d manifests the significant impact of these activities on the plasma LCFS. Specifically, following the shift from the limiter to the divertor configuration, the plasma's elongation $\kappa$, upper triangularity $\delta_\mathrm{u}$, and lower triangularity $\delta_\mathrm{l}$ surge abruptly. Fortunately, as shown in Fig. \ref{fig:dynamics model results}a, the dynamics model sensitively captures this variation. Similar robustness in predicting significant fluctuations in elongation and horizontal displacement can be observed in the initial segment of the ramp-down phase, where magnetic control still dominates despite the plasma's strong interaction with the wall producing significant non-magnetic effects. On the other hand, Fig. \ref{fig:dynamics model results}b and Fig. \ref{fig:dynamics model results}c validate the accuracy of the dynamics model to predict abrupt shape variations in both advanced divertor configuration and Internal Transport Barrier (ITB) discharges \cite{wang2024self, staebler1998theory}.

\subsection{Extrapolation Capability on Unseen Plasma Current}

\begin{figure}[!ht]
    \centering
    \includegraphics[width=\columnwidth]{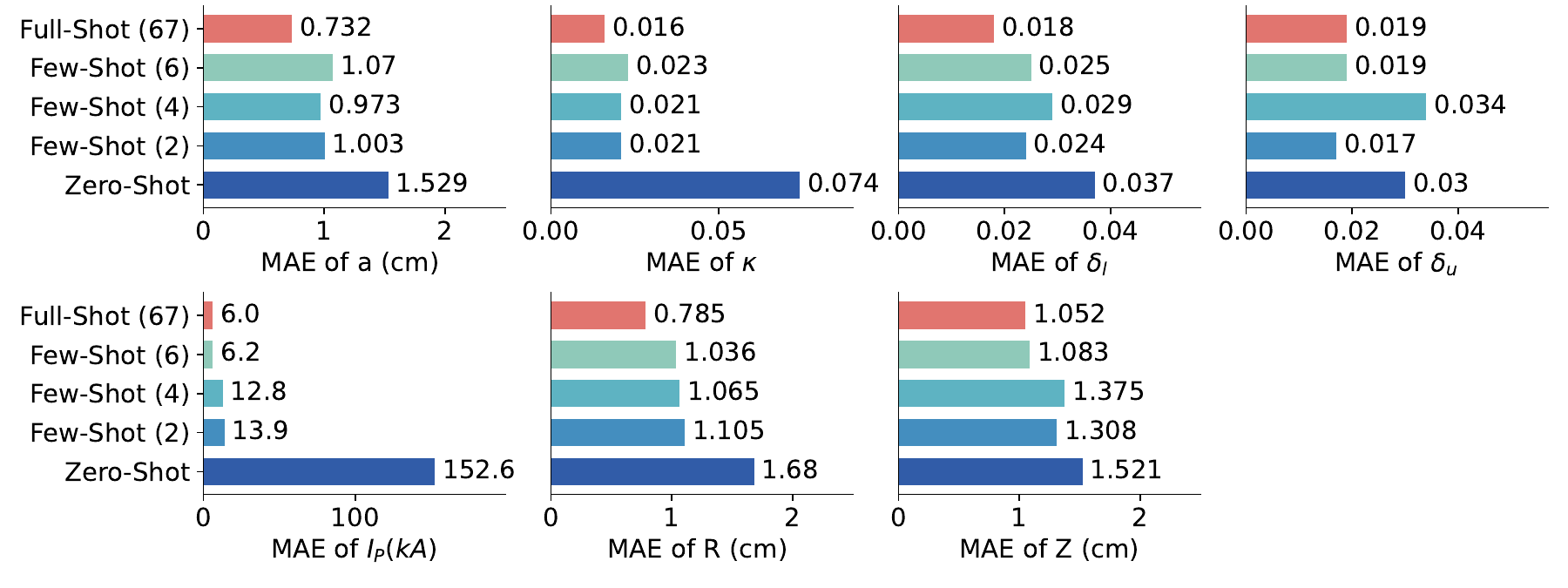}
    \caption{\textbf{Model error on $\textbf{700}$ kA discharges.} Mean Absolute Error (MAE) averaging over $30$ shots of $700$ kA discharges. The numbers in the brackets correspond to the number of $700$ kA discharges included in the training dataset.}
    \label{fig:MAE for 30 shots}
\end{figure}

\begin{figure}[!ht]
    \centering
    \includegraphics[width=.475\columnwidth]{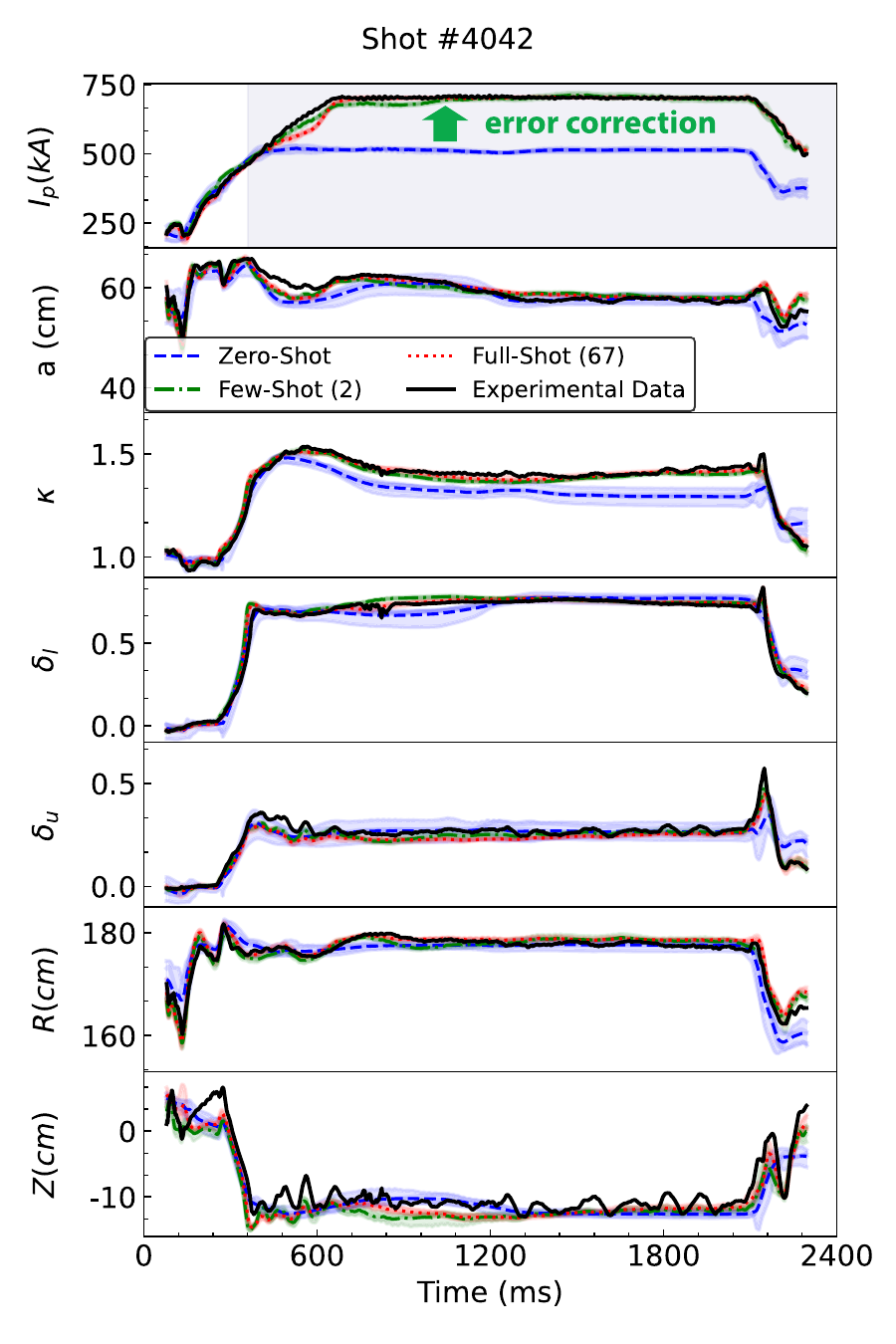}
    \caption{\textbf{Few-shot learning performance on an unseen discharge.} Test results of shot \#4042 under the dynamics model from zero-shot (blue dashed line), $2$-shot (green dashdot line), and full-shot training (red dotted line). The shaded area represents the results of the model ensemble, while actual discharge data is given in a black solid line.}
    \label{fig:4042 few shot 700}
\end{figure}
Consistent with the staged research plan in ITER to gradually increase plasma current ($I_\mathrm{p}$) \cite{iter_research-plan2024}, it is crucial to assess the extrapolation capability of a dynamics model trained on lower-$I_\mathrm{p}$ datasets when applied to higher $I_\mathrm{p}$ plasma discharge scenario. Therefore, we investigate the marginal contributions of incorporating various numbers of additional $700$ kA discharges to fine-tune a dynamics model initially trained on data with plasma currents ranging from $300$ to $600$ kA.

Fig. \ref {fig:MAE for 30 shots} presents the corresponding results averaging over a testing dataset consisting of $30$ shots of $700$ kA discharges, where the zero-shot training implies no involvement of data from $700$ kA discharges, while full-shot learning indicates using data from all available $67$ shots. As depicted in Fig. \ref{fig:4042 few shot 700}, under zero-shot training, the model gives a $152.6$ kA discrepancy and fails to reach the maximum plasma current value of $700$ kA, highlighting the generalization difficulty to higher currents without any prior information. Nevertheless, with the introduction of just $2$ or $4$ additional shots of $700$ kA data for training, the prediction accuracy improves significantly, with a $90\%$ reduction in MAE. Meanwhile, both Fig. \ref {fig:MAE for 30 shots} and Fig. \ref{fig:4042 few shot 700} indicate that with the inclusion of $6$ additional shots, the performance closely approaches that of full-shot training, demonstrating the model's strong extrapolation capability even with a limited dataset. On the other hand, for the geometric parameters in the $700$ kA discharge, a zero-shot model maintains an acceptably low prediction error, suggesting more robust extrapolative abilities for shape and position parameters. These experiments carry encouraging implications for both the predictive and control aspects of staged discharge within the upcoming ITER device.

\subsection{RL-based On-Device Control Results}
\begin{figure*}[!t] 
    \centering
    
    \includegraphics[width=1.0\textwidth]{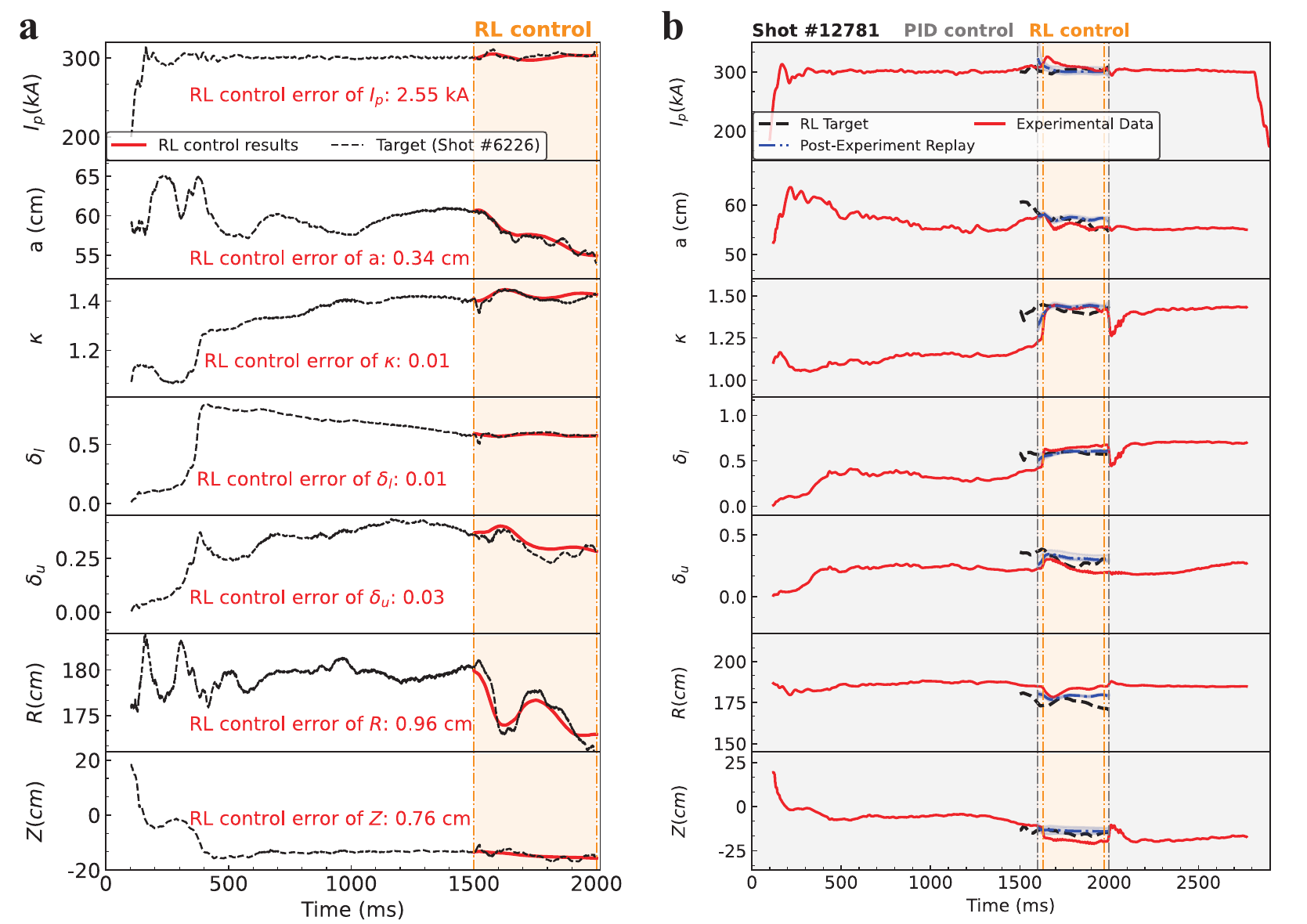}
    \caption{\textbf{Target tracking control results of RL on both simulation and actual devices.} \textbf{a}, Simulated control results of plasma current $I_\mathrm{p}$, shape and position parameters $a$, $\kappa$, $\delta_\mathrm{u}$, $\delta_\mathrm{l}$, $R$, and $Z$. The RL control results (red solid line) track the target waveform from shot \#6226 (black dashed line), with the handover time set at the $1500$ ms. \textbf{b}, in shot \#12781, the experimental data (red solid line) and the post-experiment replay of the dynamics model (blue dashdot line), as well as the target (black dashed line).}
    \label{fig:RL target tracking control results}
\end{figure*}

\begin{table}[!t] 
\centering
    \caption{Plasma Shape Parameters during RL Control for Shot \#12781} 
    \label{tab:shape_params_12781}
    \begin{tabular}{lccc}
        \toprule
        \textbf{Time (ms)} & \textbf{Elongation ($\kappa$)} & \textbf{Upper Triangularity ($\delta_\mathrm{u}$)} & \textbf{Lower Triangularity ($\delta_\mathrm{l}$)} \\
        \midrule
        1600 & 1.30 & 0.316 & 0.518 \\
        1630 & 1.40 & 0.347 & 0.612 \\
        1700 & 1.46 & 0.339 & 0.611 \\
        1800 & 1.42 & 0.269 & 0.639 \\
        1900 & 1.42 & 0.229 & 0.661 \\
        2000 & 1.38 & 0.182 & 0.620 \\
        \bottomrule
    \end{tabular}
\end{table}

In line with the long-term trajectory magnetic control in \cite{degrave2022magnetic}, we aim to learn an RL-based policy to generate direct coil voltage commands for target plasma shape. This policy is trained through interactions with an autoregressively evolving dynamics model. During training, an actor-critic-based Proximal Policy Optimization (PPO)  \cite{schulman2017proximal} is leveraged for learning to achieve the desired plasma current and shape, which is identical to the discharge result between $1500$ and $2000$ ms of shot \#6226 of the HL-3 tokamak, an H-mode discharge with ELMs and a plasma current of $300$ kA. In PPO \cite{schulman2017proximal}, the ``actor'' utilizes a policy to generate physics-reasonable magnetic coil voltage commands (see Methods) and modulate the plasma to the configuration of interest, while the ``critic'' evaluates the effectiveness of the current policy by estimating state values or advantages for potential policy refinement. Fig. \ref{fig:RL target tracking control results}a demonstrates the target tracking accuracy of the well-trained PPO actor within $1500$ - $2000$ ms' simulations. The control error of the plasma current is $2.55$ kA, or equivalently $0.85$\% of the target for a $300$ kA discharge. The control error for minor radius $a$ is $0.34$ cm, which corresponds to $0.52$\% of the $65$ cm minor radius characteristic of the HL-3 tokamak device. Additionally, the control errors for the geometric center coordinates, both $R$ and $Z$, are constrained within a maximum of $1$ cm. The results indicate that the RL agent, supported by a data-driven dynamics model, can effectively explore and assimilate strategies that are advantageous for achieving discharge control objectives.

\begin{figure}[!t]
    \centering
    \includegraphics[width=.8\columnwidth]{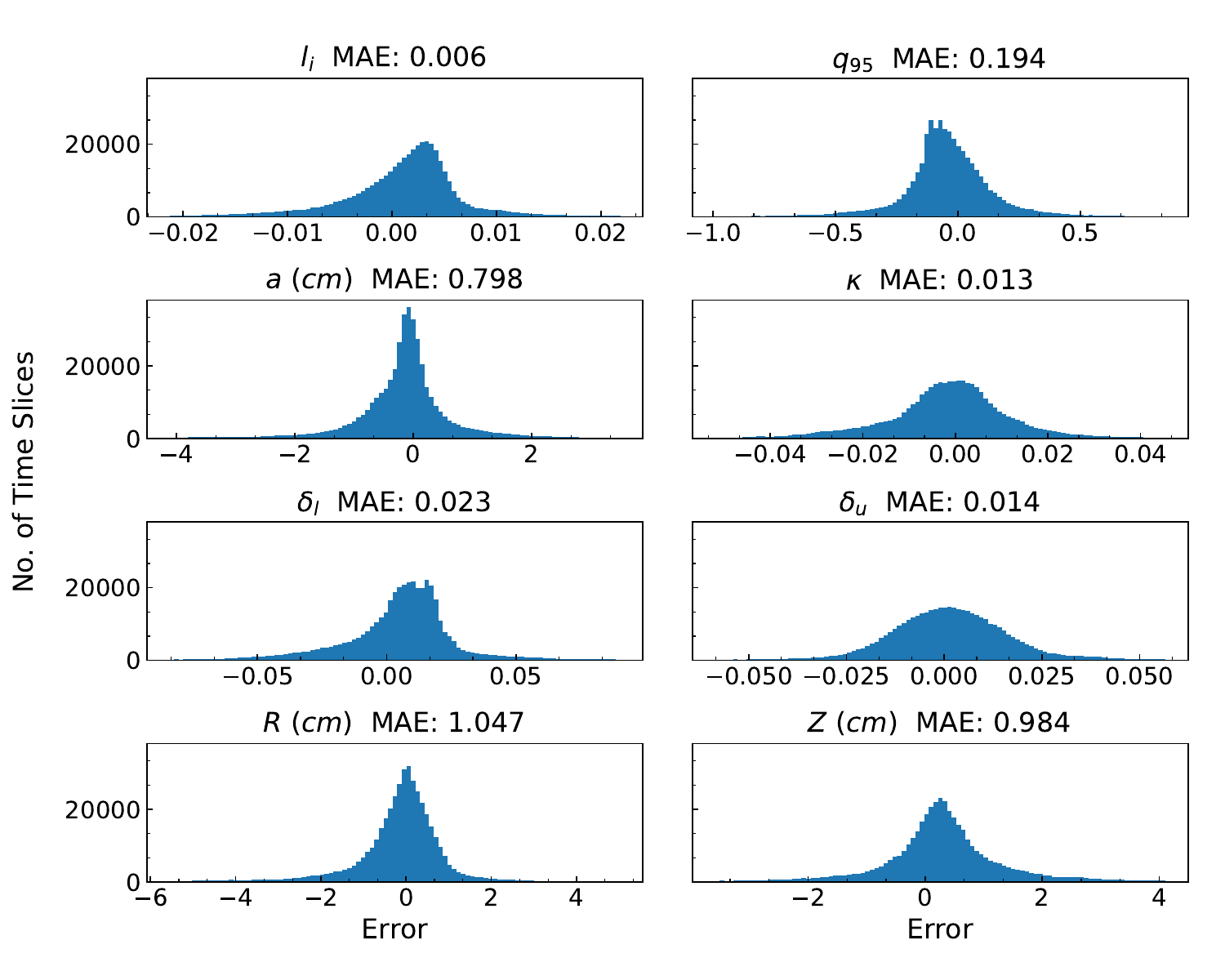}
    \vspace{-.35cm}
    \caption{\textbf{Error between EFITNN and offline EFIT.} Histograms of the prediction error between the EFITNN model and the offline EFIT code for key plasma parameters over all time slices in the test set. The Mean Absolute Error (MAE) for each parameter is shown in the title of its respective panel.}
    \label{fig:EFITNN error}
\end{figure}

Inspired by these promising results, we further evaluate the performance of the RL-based policy by integrating the well-trained PPO actor with the Plasma Control System (PCS) of the HL-3 tokamak. Consistent with \cite{degrave2022magnetic}, experiments are executed without further tuning weights of the PPO actor after training. In other words, there is a ``zero-shot'' transfer from the simulation to the real device. Meanwhile, to provide a real-time plasma shape diagnostic and complete the shape-control feedback loop, we utilize our previously developed EFITNN \cite{zheng2024real}, a surrogate model of the EFIT code \cite{lao1990equilibrium, lao1985reconstruction} consisting of several fully connected neural network layers with residual connections \cite{He_2016_CVPR}. In comparison with offline EFIT, EFITNN yields a relative error of less than $1\%$ and consumes just $0.08$ ms on NVIDIA A100 Tensor Core GPU with TensorRT. The adoption of data-driven EFITNN contributes to effectively reducing the learning expenses of the dynamics model and facilitating real-time control. A summarized comparison between EFITNN and offline EFIT is presented in Fig. \ref{fig:EFITNN error}.

\subsubsection{RL-based $400$-ms Magnetic Control for Target Tracking}
\label{sec:rl_exp_1}
To thoroughly evaluate the flexibility of RL-based target tracking, we adopt a progressive approach. Initially, we assess the agent's target tracking ability and ensure it aligns with the simulations. For this purpose, the experimental target remains identical to the reference shot \#6226 for PPO training and the PPO actor responds to the variations in plasma shape and position parameters diagnosed by EFITNN. Despite the consistency in targets, noticeable differences arise during the PID control due to the inherent uncertainties in tokamak discharge and simultaneously coupled operations for HL-3 maintenance and testing.
A series of RL control experiments have been performed on {HL-3}.

Typically, in shot \#12781, alongside the traditional PID control, we conduct an RL-based magnetic control from $1600$ ms to $2000$ ms with $30$ ms transition periods at the start and end. 
Fig. \ref{fig:RL target tracking control results}b provides the corresponding control results and Table \ref{tab:shape_params_12781} details the evolution of elongation ($\kappa$) and upper and lower triangularities ($\delta_\mathrm{u}$ and $\delta_\mathrm{l}$) during the RL control phase.
During its operational phase, the RL controller demonstrates excellent tracking performance for the four plasma shape parameters ($a$, $\kappa$, $\delta_\mathrm{u}$, and $\delta_\mathrm{l}$). It rapidly adjusts these parameters to their target values and subsequently maintains their stability with high precision. On par with PID, the controller capably maintains the radial ($R$) and vertical ($Z$) positions within a stable range, with precision slightly lower than that of shape parameters.  
Additionally, the control handover induces a brief fluctuation in the plasma current ($I_\mathrm{p}$), which the RL controller promptly corrects, restoring it to the target level.
The experimental results confirm that the RL controller possesses the capability for precise plasma shape regulation. In contrast to the conventional PID system, which is limited to controlling plasma positions ($R$ and $Z$) and current ($I_\mathrm{p}$), the RL controller successfully extends the closed-loop scope to the simultaneous regulation of four additional key shape parameters ($a$, $\kappa$, $\delta_\mathrm{u}$, and $\delta_\mathrm{l}$).

\subsubsection{RL-based Control for Zero-Shot Adaptation to Changed Triangularity Target}
\label{sec:rl_exp_2}
\begin{figure*}[!tb]
\centering
    \includegraphics[width=0.6\textwidth]{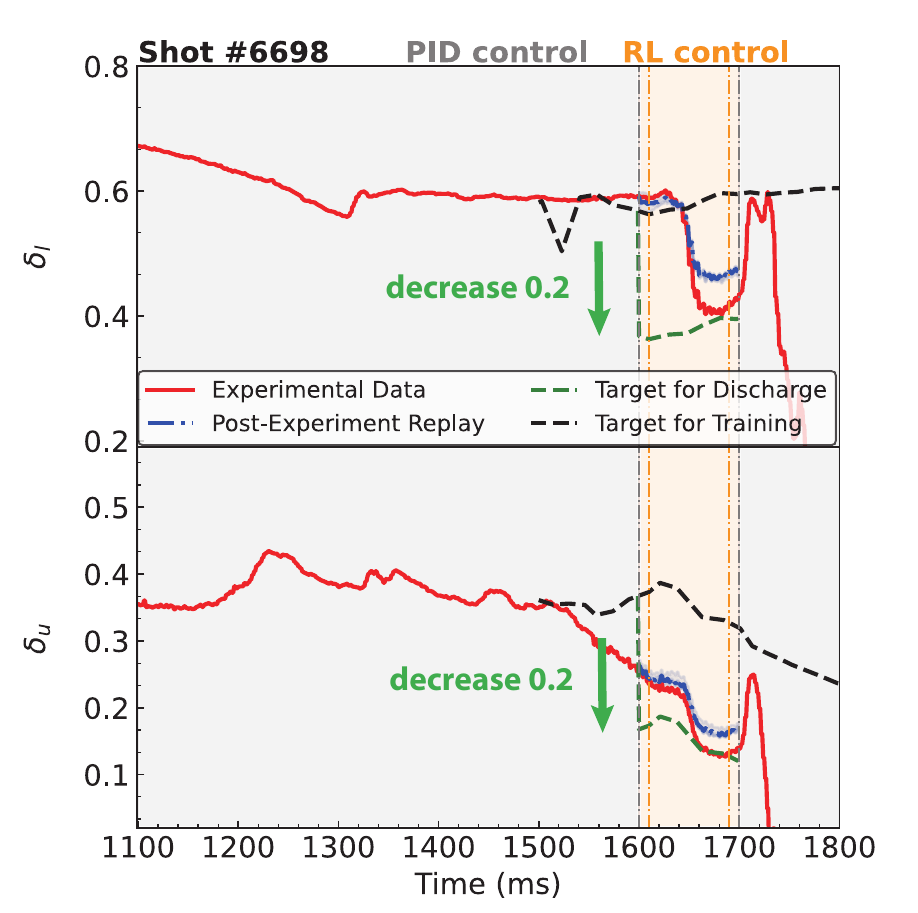}
    \caption{\textbf{Zero-shot triangularity control in shot \#6698.} The upper and lower triangular control results of shot \#6698. Compared to the target used for RL training (black dashed line), the target for this discharge (green dashed line) has a triangularity decreased by $0.2$. The red solid line illustrates the experimental result, and the blue dashdot line represents the post-experiment replay of our dynamics model.}
    \label{fig:RL delta control results}
\end{figure*}

\begin{table}[!t]
\centering
    \caption{Triangularity Parameters during RL Control for Shot \#6698}
    \label{tab:shape_params_6698}
    \begin{tabular}{lcc}
        \toprule
        \textbf{Time (ms)} & \textbf{Upper Triangularity ($\delta_\mathrm{u}$)} & \textbf{Lower Triangularity ($\delta_\mathrm{l}$)} \\
        \midrule
        1610 & 0.235 & 0.589 \\
        1630 & 0.224 & 0.597 \\
        1640 & 0.220 & 0.577 \\
        1650 & 0.189 & 0.482 \\
        1660 & 0.144 & 0.412 \\
        1690 & 0.131 & 0.423 \\
        \midrule
        \textbf{Target} & \textbf{0.128} & \textbf{0.396} \\
        \bottomrule
    \end{tabular}
\end{table}
Triangularity is a critical shape parameter in tokamak plasmas, significantly affecting both vertical stability and the structure of the plasma boundary \cite{schuster2006role}. Consequently, the ability to control triangularity is crucial for optimizing tokamak plasma equilibria and performance. To validate the flexibility of RL, we intentionally decrease the upper and lower triangular shape parameters of the plasma by $0.2$ from those target values during the training phase. The performance of the PPO actor on triangularity targets not encountered during training allows us to evaluate the agent's adaptability to varying control objectives. 

In shot \#6698, similar to the procedure described in the preceding subsection on target tracking control, RL assumes the discharge control task from PID from $1600$ ms to $1700$ ms with $10$ ms transition periods at the start and end. Fig. \ref{fig:RL delta control results} and Table \ref{tab:shape_params_6698} illustrate the corresponding control results of the upper and lower triangularity. Both the lower triangularity $\delta_\mathrm{l}$ and the upper triangularity $\delta_\mathrm{u}$ successfully reach the new target values. Specifically, the lower triangularity $\delta_\mathrm{l}$ swiftly progresses towards the target following a transition period of approximately $10$ ms, finally reaching it around $1650$ ms. 
Meanwhile, the upper triangularity $\delta_\mathrm{u}$ follows a gradual trajectory toward the target during the initial $10$-ms transition phase, accelerates its adjustment after around $20$ ms, and aligns with the target at approximately $1650$ ms. It then maintains alignment with the target trajectory for the remaining $50$ ms. 
These zero-shot triangularity control results verify the versatility of RL towards effective control in previously unencountered scenarios.
Furthermore, the post-experiment replay of the dynamics model from $1600$ ms to $1700$ ms reveals a consistent trend, underscoring the model's generalizability and robustness for supporting RL in exploratory applications.

\section{Conclusions}
We present a technique to learn a fully data-driven dynamics model to characterize the evolution of plasma in a tokamak device. The dynamics model, which leverages multiple components to mitigate the compounding error issue and improve accuracy during autoregression, exhibits high fidelity and appealing extrapolation capability. It effectively supports the fast training of an RL-based high-dimensional trajectory control policy, which together with the real-time plasma diagnostic empowered by EFITNN, can directly manipulate the magnetic coil voltages to transform the plasma state toward desired configurations. 

\begin{figure}[!t]
    \centering
    \includegraphics[width=.9\textwidth]{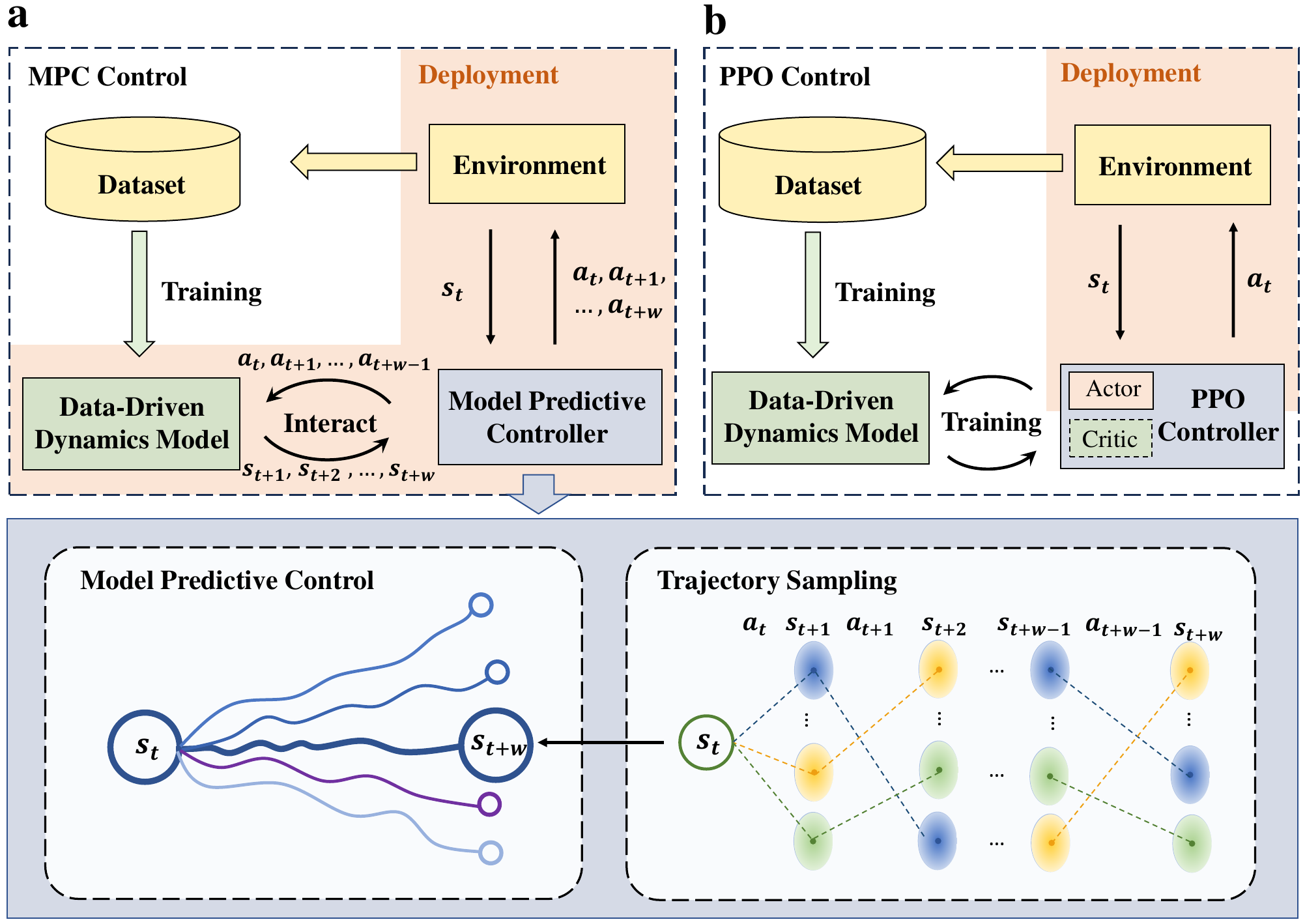}
    \caption{\textbf{Comparison of the training and deployment between MPC and PPO.} \textbf{a,} MPC with online, limited-term dynamics model autoregression and \textbf{b}, PPO interacting with dynamics model for training only.}
    \label{fig:MPC}
\end{figure}

Our work demonstrates the feasibility of data-driven, end-to-end learning and control of plasma current and shape. By avoiding reliance on first-principle-based simulators \cite{degrave2022magnetic} and auxiliary controlling commands from historical data or controllers \cite{seo_development_2022,seo2021feedforward,seo_avoiding_2024,char2023offline, char2024pid,abbate_general_2023}, our approach significantly differs from existing works. 
High-dimensional control over a learned model falls into the general scope of model-based reinforcement learning (MBRL) \cite{moerland_modelbased_2023}. In this regard, as in Supplementary Figure 1a, instead of evolving the plasma state, Ref. \cite{seo_avoiding_2024} mainly uses the output of the learned dynamics model to generate a competent reward function, which limits its applicability to localized reactive control only. Meanwhile, as in Fig. \ref{fig:MPC}, Ref. \cite{abbate_general_2023} conducts a receding (limited-term) horizon control through computation-intensive MPC, which involves cumbersome online autoregression of the dynamics model for trajectory sampling, making real-time deployment challenging \cite{Lore_2023}.
On the contrary, as in Supplementary Figure 1b, our work provides a proof-of-concept study on long-term high-dimensional dynamics model learning, which further enables the training and deployment of the independent PPO actor for complex magnetic control. 

Our research presents a promising direction for routine discharge operations in future tokamak devices.
In the future, we will focus on further improving the overall performance and generalization capability of the control model.  
Building upon this foundation, we aim to address more challenging control tasks, including the intelligent avoidance of plasma instabilities and disruptions \cite{murari2024control, sheikh2018disruption, barr2021development}. Moreover, we plan to conduct transfer learning across tokamak devices \cite{murari_control_2024}, further verifying the potential advantages of a fully data-driven method over first-principle-based simulators.

\renewcommand{\figurename}{Extended Data Fig.}
\renewcommand{\tablename}{Extended Data Table}
\setcounter{figure}{0}
\setcounter{table}{0} 
\section{Methods}
\subsection{HL-3}
The HL-3 (previously named HL-2M) Tokamak is an experimental fusion device constructed at the Southwestern Institute of Physics in China. It is designed to operate at plasma current $I_\mathrm{p}$ =$3$ MA, toroidal field $B_\mathrm{t}=3$ T, major radius $R=1.78$ m, minor radius $a=0.65$ m, elongation $\kappa \leq 1.8$, triangularity $\delta > 0.5$ \cite{zhong2024china}. HL-3 has recently focused on high-current, high $\beta_\mathrm{N}$ discharges to achieve high performance plasma, establishing critical foundations for burning plasma research. 
As shown in Fig. \ref{fig:diagram}a, the poloidal field configuration of HL-3 is intricately managed by a Central Solenoid (CS) coil, complemented by an ensemble of eight pairs of Poloidal Field (PF) coils strategically distributed as inner coils (PF1-PF4), top-bottom coils (PF5-PF6), and outer coils (PF7-PF8), with the latter trio playing a pivotal role in establishing the divertor configuration, hence their nomenclature as divertor coils \cite{song2021plasma, song2019first, xue2019integrated}. In particular, HL-3 directly manipulates the voltage of $17$ coils, comprising of $1$ CS coil and $16$ PF coils. This manipulation indirectly influences the coil currents and regulates the magnetic field within the vacuum chamber, enabling the confinement and discharge of plasma.

\subsection{Data Pre-Processing}
Typically, the plasma discharge process can be divided into three distinct phases (i.e., ramp-up, flat-top, and ramp-down phases). However, for the ramp-up phase's plasma, the relatively low current and the difficulty in calculating the eddy currents in passive conductor structures (such as the vacuum chamber and the first wall) often result in inaccurate shape calculations, weakening the correlation between the coil voltage commands and plasma shape. Therefore, considering that the primary objective of the dynamics model is to capture the complex evolution of the plasma state and ultimately facilitate efficient RL training, model training is exclusively based on data from the ramp-up phase where the plasma current surpasses $100$ kA and flat-top phase with a duration of at least $500$ ms. Based on the historical discharge logs of HL-3 across campaigns from 2022 to 2024, after such a meticulous selection procedure, a dataset comprising $832$ successful discharges has been curated, with a temporal resolution of $1$ ms, totaling $1,833,516$ time slices. The categorization of these $832$ shots by flat-top plasma current is detailed in Supplementary Table 1.

Acknowledging the importance of the historical plasma state and coil commands, the dynamics model integrates a time series as its input to better capture the plasma evolution towards equilibrium. Specifically, given the $1$-kHz control frequency, the model uses a $30$-ms historical dataset for single-step prediction of plasma current, shape and position parameters, as well as the coil currents. A summarized list of the input and output variables of the dynamics model can be found in Supplementary Table 2. Briefly, this model learns up to $25$ variables related to plasma current, shape, and position, as well as coil current from a $44$-dimensional input, which includes autoregressive predictions of these $25$ variables, $17$ executed coil voltage configurations, the toroidal magnetic field, and loop voltage. In addition, $n = 30$ represents the length of the time series, and $\Delta{t} = 1$ ms is the temporal resolution. As the CS coil is composed of two parallel connected sets of coils, the dimension of coil current ($I_\mathrm{c}$) equals that of coil voltage ($U$) plus $1$. 

Throughout the discharge experiments in HL-3,  engineering conditions change continuously. As a result, we utilize the most recent discharges for testing while earlier datasets are used for training. This methodology contributes to learning from updated data distributions and accommodates to latest discharge environment. To prevent model overfitting, we designate $50$ discharges from shot \#5893 to shot \#6055 as the validation set, and $70$ discharges, including shots \#6056-\#6265 and \#5095-\#5105, as the test set. Note that the range from shot \#5095 to \#5105 corresponds to $10$ discharges of advanced divertor configuration, which exhibit more pronounced morphological variations than standard discharges. During training, a $15$-ms smoothing process is implemented on the training data to effectively mitigate non-electromagnetic interference. Since the training dataset suffers from a paucity of similar data instances, the corresponding results in Supplementary Figure 4 could partially validate the accuracy and extrapolation capability of the dynamics model. 

\subsection{DNN Design}
\subsubsection{DNN Structure for Dynamics Model}

Due to the practical demands, plasma discharges are often carried out with identical or similar parameter sets to replicate specific physical outcomes, resulting in a lack of diversity in the dataset. Therefore, although the Transformer excels in parallel computation and can apprehend holistic information owing to its self-attention mechanism \cite{vaswani2017attention}, it faces overfitting issues under limited and narrow data distribution. The testing MAE for both the Transformer and LSTM networks, as shown in Supplementary Figure 2, suggests that under the same training methods, dynamics models utilizing LSTM networks typically outperform those based on the Transformer. As a result, we opt for LSTM as the backbone of our dynamics model.

We utilize a stack of DNNs including a three-layer LSTM network, a self-attention module, and a three-layer fully connected MultiLayer Perceptron (MLP). Specifically, each LSTM layer comprises $128$ hidden neurons with a dropout rate of $0.2$. For the $30$ consecutive steps' $44$-dimensional input, as in Supplementary Table 2, the third LSTM layer produces an intermediate hidden layer output with dimensions of $30\times 128$. Subsequently, the inner product between each $128$-length tensor and the last one is computed, followed by a softmax-based normalization operation to scale the values between $0$ and $1$. These self-attention values serve as attention weights to aggregate the original LSTM outputs as a $128$-length vector. The integration of a self-attention mechanism into the LSTM allows the assignment of distinctive weights to the intermediate output, rather than placing equal emphasis and neglecting the inter-differences. Fig. \ref{fig:trick_all}a validates the effectiveness of this modification. Afterward, the self-attention-processed vector undergoes three MLPs with $128$, $64$, and $25$ neurons and a ReLU activation function, culminating in the single-step prediction of $25$ variables as specified in Supplementary Table 2.

\subsubsection{DNN Structure for EFITNN}
EFITNN \cite{zheng2024real} is a surrogate model trained with input and output provided by an offline Equilibrium FIT (EFIT) code \cite{lao1990equilibrium, lao1985reconstruction}. It accepts input signals from poloidal array of pick-up coils, flux loops, and coil currents, and concatenates them into a $68$-element vector. EFITNN supports the prediction of $8$ scalar values including six plasma shape parameters as well as safe factor in 95\% minor radius $q_\mathrm{95}$ and plasma inductance $l_\mathrm{i}$. Besides, it can produce predictions for poloidal magnetic flux distribution $\Psi_\mathrm{rz}$ and toroidal current density distribution $J_\mathrm{p}$ with a resolution up to $129\times 129$. For these heterogeneous outputs, EFITNN adopts a shared $4$-layer MLPs. Subsequently, for scalar prediction, another $4$-layer MLPs with residual connections \cite{He_2016_CVPR} are leveraged, while $\Psi_\text{rz}$ and $J_p$ relies on deconvolutional networks \cite{Zeiler2014Visualizing}. Note that each MLP layer consists of $512$ neutrons with a GeLU activation function. EFITNN adopts simultaneous training for this multi-task learning, and we kindly refer to \cite{zheng2024real} for the training details. Fig. \ref{fig:EFITNN error} manifests the prediction accuracy after training. In addition, if only plasma shape and position parameters are predicted, EFITNN, without the use of deconvolutional networks, consumes $0.08$ ms on NVIDIA A100 Tensor Core GPU with TensorRT, fully supporting the real-time deployment.

\subsubsection{DNN Structure for PPO}
We train the RL agent using PPO \cite{schulman2017proximal}, which, as an actor-critic framework, updates both the policy and value networks through the interactions with the well-trained dynamics model. In our experiments, both networks are structured as three-layer MLPs. The policy network uses a tanh activation function and consists of $64$ neurons per layer, while the value network employs a ReLU activation function and comprises $128$ neurons per layer.

\subsection{Training of High-Fidelity Dynamics Models}
\subsubsection{Scheduled Sampling}

Without any preventive training measures, the autoregressive nature of dynamics models leads to a gradually increasing deviation between the actual data and model predictions due to compounding error, as illustrated in Supplementary Figure 3. Directly applying such a model in practice can result in exposure bias \cite{schmidt2019generalization}, incurring reduced efficacy and severe misalignment between the training and testing modalities. Therefore, during our training, we incorporate scheduled sampling \cite{bengio2015scheduled, vlachas2023learning} to stochastically determine the input, by selecting between the model's past predictive outputs and the current-step actual data according to a teacher-forcing ratio $p \in [0,1]$. In other words, as the ratio $p$ approaches $1$, a greater proportion of the real input data will be used. Conversely, as $p$ turns to $0$, the model transitions to a strictly autoregressive mode, solely relying on the preceding steps' predictions. During the training, we initialize the ratio $p$ to $1$ and linearly decay it to $0$ over $300$ epochs. This approach ensures rapid convergence during training and facilitates a seamless shift towards an autoregressive scheme for training RL. 

In contrast, \cite{seo2021feedforward} adopts a different approach to intentionally add a zero-mean Gaussian noise layer to the training data. Nevertheless, apart from the conflict with the data smoothing operation during training, the comparison in Supplementary Figure 4 also suggests its inferiority to scheduled sampling. Specifically, the diverse impact on high-dimensional multi-task learning makes selecting the appropriate variance of added Gaussian noise challenging. For example, with an amplitude of $0.1\sigma_{x}$, where $\sigma_{x}$ indicates the standard deviation of the input data, the model exhibits the lowest error in predicting $\kappa$ but the highest error in predicting $I_\mathrm{p}$. Conversely, the incorporation of scheduled sampling maintains stability in significantly reducing prediction errors for $25$ variables. Therefore, we primarily take account of scheduled sampling in this paper.

\subsubsection{Adaptive Weight Adjustment}
To unanimously provide multi-task predictions for $25$ variables, in Supplementary Table 2, the calibration of relative weights assigned to the predictive losses of individual variables could heavily affect the model performance. Mathematically, for a space $\mathcal{V}$ containing $25$ variables, the aggregated loss function given the $44$-dimensional input $\boldsymbol{x}$ shall be a scalar $\mathrm{L}_{\mathcal{V}}\left(\boldsymbol{y}_{\mathcal{V}}, \hat{\boldsymbol{y}}_{\mathcal{V}};\boldsymbol{x} \right)$ computed from $25$ individual loss functions $L_{v}\left(y_v, \hat{y}_v;\boldsymbol{x}\right)$ weighted by $c_v$, where $y_v$ and $\hat{y}_v$ correspond to actual data and model prediction for the variable $v \in \mathcal{V}$, respectively while $\boldsymbol{y}_{\mathcal{V}}$ and $\hat{\boldsymbol{y}}_{\mathcal{V}}$ denote the concatenation of $y_v$ and $\hat{y}_v$, $\forall v \in \mathcal{V}$. 
Due to the discrepancies in distribution and noise among the variables, a straightforward averaging of individual loss functions can adversely impact the model's accuracy. Meanwhile, manual weight tuning is an arduous and time-intensive effort. Therefore, we utilize an adaptive weight adjustment technique by regarding the loss weight $c_v$ as a trainable single variate\cite{kendall2018multi}. To prevent the loss weights from taking negative values during the optimization process, which could lead to negative loss values, we also employ a non-negative regularization constraint \cite{liebel2018auxiliary} to enhance the learning robustness. To sum up, the final loss function for training can be calculated as
\begin{equation}
\label{eq:loss_function}
\mathrm{L}_{\mathcal{V}}\left(\boldsymbol{y}_{\mathcal{V}}, \hat{\boldsymbol{y}}_{\mathcal{V}};\boldsymbol{x} \right) =  \sum_{v \in \mathcal{V}} \left[ \frac{1}{2 \cdot c_v^2} \cdot L_{v}\left(y_v, \hat{y}_v;\boldsymbol{x}\right)  +\ln \left(1+c_v^2\right) \right].
\end{equation}

\subsubsection{Training Details}
In training the dynamics model, we initially segment all the discharge data of training shots into consecutive groups with a fixed length of $30$ ms, followed by a shuffling process on all the segmented data. We use DataLoader, a PyTorch library, to load the processed training data, with a batch size set as $4,096$. In each training round, we randomly utilize a batch of data without replacement for training and an epoch completes until all groups have been utilized. We employ the Adam optimizer for optimizing network parameters, with an initial learning rate of $0.001$ for both the dynamics model and the adaptive weight adjustment. To prevent training instability due to an excessively high learning rate in the later stages of training, we adjust the learning rate using an exponential decay method. Specifically, we reduce the learning rate by a decay factor of $0.95$ every $5$ epochs, over a total of $300$ training epochs.

Finally, in addition to Fig. \ref{fig:trick_all}, Supplementary Figure 5 shows the accuracy for predicting all the $18$ coil currents.

\subsection{Training and Deployment of RL-based Magnetic Controller}
\subsubsection{State, Action, and Reward of RL}
In this study, we utilize RL to control the plasma current ($I_\mathrm{p}$) and six shape and position parameters ($a$, $\kappa$, $\delta_\mathrm{l}$, $\delta_\mathrm{u}$, $R$, $Z$), and denote the concatenation of current observations as $\mathbf{o}$. To ensure the integrity and stability of HL-3 plasma discharges, we select the actual waveform of shot \#6226, an H-mode discharge with ELMs and a plasma current of $300$ kA, as the desired target $\mathbf{d}$ for control. As shown in Supplementary Table 3, we consider a $14$-dimensional state  $\mathbf{s}$ by amalgamating the observations $\mathbf{o}$ and desired target $\mathbf{d}$, while the $17$-dimensional action $\mathbf{a}$ encompasses voltage commands of $1$ CS coil and $16$ PF coils. To ensure control safety and hardware integrity, we impose operational constraints by stipulating that the dynamic rate of change between two consecutive voltage commands must not exceed $20~\mathrm{V}~\mathrm{ms}^{-1}$. In other words, for any coil, any voltage command that changes beyond this limit will be clipped to adhere to the safety constraints.

Typically, RL relies on explicit targets for reward function design. In fusion experiments, however, diverse targets from varied discharge templates complicate reward engineering and can undermine model efficacy.
Therefore, we devise the reward function $r(\boldsymbol{s},\boldsymbol{a})$ from two orthogonal perspectives. First, consistent with \cite{degrave2022magnetic,TRACEY2024114161}, the reward function primarily takes account of the element-wise divergence between the current observation $\boldsymbol{o} \triangleq [o^{(1)},\cdots, o^{(7)}]$ and desired target $\boldsymbol{d} \triangleq [d^{(1)},\cdots, d^{(7)}]$. To merge the divergence components into a scalar reward and emphasize those control variables with sub-optimal control results, a smooth-max function \cite{TRACEY2024114161} is leveraged. Second, owing to the constraints of the power module, the coil current is permitted to cross zero only once. Therefore, contingent on the coil current predicted by the dynamics model, we impose a penalty term to inhibit the agent from exploring control commands that would result in multiple zero crossings of the coil current. To sum up, the reward for a state-action pair $(\boldsymbol{s},\boldsymbol{a})$ is given as
\begin{equation}
\label{eq:reward_func}
r(\boldsymbol{s},\boldsymbol{a};\boldsymbol{\omega},\alpha) = -\frac{\sum_{i=1}^{7} \omega_i \vert o^{(i)} - d^{(i)} \vert  \exp(-\alpha \vert o^{(i)} - d^{(i)} \vert)}{\sum_{i=1}^{7} \omega_i \exp(-\alpha \vert o^{(i)} - d^{(i)} \vert)} + \sum\nolimits_{i=1}^{17} p_i,
\end{equation}
where weights $\boldsymbol{\omega} \triangleq [\omega_1, \cdots, \omega_7]$ in the softmax function provide the (relative) importance of each component and are initialized as $[3, 2, 2, 3, 1, 2, 2]$. The modulation factor $\alpha$, which affects the trade-offs between ``easy'' and ``hard'' (to satisfy) components \cite{TRACEY2024114161}, is set as $-1$ in our experiment. Besides, for $17$ coils, the penalty term $p_i$ for coil $i$ is designed as $-1$ when the coil current exceeds one zero-crossing while nulls otherwise. 

\subsubsection{Dynamics Model-based Training}
We utilize the trained dynamics model as the interaction environment, through which the RL agent explores suitable actions to achieve the desired plasma current and shape. Specifically, as the HL-3 tokamak is a preliminarily established device and initiates the global open campaigns in 2024 and 2025, a significant number of discharge shots have been scheduled for international joint physical experiments. Consequently, the proposal for RL-based control has to share discharges with other experiments. 
The choice of shot \#6226 for target tracking is also motivated by the potential to maximize RL’s takeover opportunities, considering the frequent use of similar shot templates in planned experiments on the HL-3 tokamak. As highlighted in Fig. \ref{fig:RL delta control results}, the promising outcomes underscore the inherent generalization capability, which suggests robust applicability to a variety of other discharge templates.
In addition, given the potential uncertainty in freshly deploying RL-based control on HL-3, the early phase of a discharge, prior to $1500$ ms, is primarily allocated to other physics experiments. Therefore, during training, RL assumes the $1$-kHz control since the $1500$ ms of the discharge, and the initial observations $\boldsymbol{o}$ are synchronized with those of shot \#6226, ensuring a smooth transition between PID control and RL control. 
Furthermore, we take episodic training from $1500$ ms to $2000$ ms of the discharge. Without loss of generality, at each time step $t$, the RL agent determines an action $\mathbf{a}_t$ according to the state $\mathbf{s}_{t}$, while the dynamics model yields next-step plasma current as well as the six shape and position parameters $\mathbf{o}_{t+1}$ based on $\mathbf{s}_{t}$ and $\mathbf{a}_t$. Note that given the slow-changing nature of the toroidal magnetic field and loop voltage, these two values are considered invariant and maintain the values same as the handover time of shot \#6226.  Hence, a reward $r(\mathbf{s}_{t},\mathbf{a}_t)$ can be computed as in Equation \eqref{eq:reward_func}. Finally, the episodic reward is derived from the discounted accumulative reward with a discount factor $\gamma$ set as $0.98$.
To strike a balance between the variance and bias in the value function estimation, we employ the Generalized Advantage Estimation (GAE) \cite{fujimoto2018addressing} to compute the advantage function. In addition, other parameters for the PPO are set in a standard way. Specifically, we set the entropy coefficient, the clipping parameter $\epsilon$, and the $\lambda$ parameter for GAE as $0.05$, $0.2$, and $0.95$, respectively.
Following this, the PPO algorithm \cite{schulman2017proximal} can iteratively update the agent's actor and critic networks. Fig. \ref{fig:RL target tracking control results}a illustrates the tracking accuracy of the trained agent toward the target waveform within the simulation setting and suggests that the RL agent, supported by a data-driven dynamics model, has effectively explored and assimilated strategies that are beneficial for achieving discharge control objectives.

The aforementioned single-template training necessitates retraining the model with a new target for changed operational scenarios (e.g., targeting a $700$ kA plasma current). Despite this requirement for retraining, the process is efficient. Systematic evaluations across different GPU platforms (Supplementary Figure 9 and Supplementary Table 4) substantiate this efficiency. The PPO-based controller converges within $500$ training episodes ($25,000$ interaction steps). Utilizing a standard configuration (PyTorch 2.1.0, CUDA 12.2, batch size $512$), training on an NVIDIA GeForce RTX 4090 completes in just $22$ minutes. Such rapid training underscores the framework's practical adaptability, even with the need for template-specific retraining.
\subsubsection{On-Device Deployment}
To ascertain the practicability of the trained RL agent as well as the accuracy of the dynamics model, we integrate the PPO actor into the PCS of HL-3 to undertake RL-based control in real discharges. Since the PCS operates under the Redhat $6.9$ system and performs real-time computations in C language, we convert the Pytorch-based PPO actor into a compatible ONNX (Open Neural Network Exchange) \cite{jin2020compiling} format, which offers cross-platform deployment capability and high execution efficiency. During real discharges, the RL agent receives plasma shape and position parameters from EFITNN  \cite{zheng2024real} while the plasma current is measured by the CODIS-RTC system via reflective memory in real time. The agent then generates coil voltage commands and delivers them to actuators through the PCS, completing a closed-loop control period of $1$ ms, synchronized with that for the PID control. 

As mentioned earlier, the open campaigns on HL-3 have attracted a considerable number of international joint experiments, limiting the opportunities for testifying RL-based control. Eventually, in a shared discharge, the handover to RL-based control has to be delayed until other physics experiments involving auxiliary heating are completed. To avoid any potential plasma disturbances caused by shutting down the auxiliary heating at $1500$ ms, which could impact the analysis of experimental results, we have further postponed the start of RL-based control to $1600$ ms. In other words, we activate the RL control at $1600$ ms and revert to the PID control after a predefined window. To ensure stability incurred during these mode switches, we establish a transition period, during which a weighted sum of commands from both modes is undertaken and the weight of superseding control mode linearly increases. 

Our initial RL control experiments are conducted during the \#6691-\#6698 shot series on HL-3. Due to the coincidence with the development of the fast vertical position control system, the experiments undergo severely unstable discharge conditions, with only three effective discharges (i.e., shots \#6691 (Supplementary Figure 6a) , \#6693 (Supplementary Figure 7) and \#6698 (Supplementary Figure 6b)) sustaining the flat-top phase beyond $1600$ ms. Furthermore, to ensure operational safety, the RL control window is limited to $100$ ms, with $10$ ms transition periods at its start and end.
Unfortunately, the conventional PID and fast vertical position control still exhibits limited capability to suppress emerging voltage variations after reverting from RL to PID control, frequently resulting in plasma disruptions. 
More recently, following the maturation of the fast vertical position control system for routine operations, we replicate these experiments during the \#12777-\#12781 shot series, by adopting the same actor model. This second campaign yields three effective RL-controlled discharges: shots \#12777 ($100$ ms), \#12779 ($260$ ms), and \#12781 ($400$ ms). For these shots, $30$ ms transition periods are implemented at the start and end. All three effective discharges are completed with no disruptive events. Fig. \ref{fig:RL target tracking control results}b illustrates the complete discharge for shot \#12781. The complete discharges for the remaining five successful RL-controlled shots (from both experimental campaigns) are presented in Supplementary Figures 6, 7, and 8. To quantitatively assess the performance differences between RL–based control and fixed voltage control, we conduct a comparative experiment using shot \#12777. For shot \#12777, it adopts an RL control solely until $1700$ ms without the last $30$-ms transition period, while from $1700$ ms to $1730$ ms, the control voltages are deliberately fixed at their $1700$ ms values. As depicted in Supplementary Figure 8, the resulting oscillations in plasma current and shape confirm that, without closed-loop control, plasma stability is significantly compromised, demonstrating the critical need for an effective closed-loop control. Supplementary Figure 10 illustrates the plasma current for the seven shots from these campaigns in which RL control is not executed.

\subsubsection{Post-Experiment Analyses}
As illustrated in Fig. \ref{fig:RL target tracking control results}b and Fig. \ref{fig:RL delta control results}a, the post-experiment replay for shot \#12781 and shot \#6698 is calculated using the dynamics model. Specifically, the dynamics model starts to evolve according to the input variables in Supplementary Table 2, set as practical values at the handover time, as well as the practically executed coil voltage commands, which are generated by RL during on-device deployment. In other words, such a post-experiment replay represents the autoregressive output of the dynamics model, and its closeness to the EFITNN-based diagnostic measurement contributes to verifying the accuracy of the dynamic model. 

On the other hand, as shown in Fig. \ref{fig:dynamics model results}d, to visually display the prediction and control effect on the plasma shape and position, we calculate the four extreme points of the cross-section based on the six shape and position parameters obtained from the EFITNN computation. The plasma boundary is then approximated as the connection of four-quarter ellipses \cite{dong2024adapted}.

To better understand the control logic of the RL controller, we perform a more in-depth analysis of the experimental results from shot \#$12781$. As illustrated in Fig. \ref{fig:RL target tracking control results}b, while the controller demonstrates high-fidelity tracking of the primary shape parameters ($a$, $\kappa$, $\delta_\mathrm{u}$ and $\delta_\mathrm{l}$), the positional parameters ($R$ and $Z$) exhibit persistent steady-state deviations from their targets. This differential control efficacy is primarily attributed to three factors: (1) the inherent strong coupling among the shape-defining parameters ($a$, $\kappa$, $\delta_\mathrm{l}$ and $\delta_\mathrm{u}$), which are co-regulated by the PF coils; (2) the non-negligible influence of non-electromagnetic experimental disturbances, such as gas puffing and auxiliary heating, which could significantly affect the evolution of $R$ and $Z$; and (3) an implicit control priority mechanism within the RL policy -- the policy prioritizes the correction of $\kappa$, $\delta_\mathrm{l}$, $\delta_\mathrm{u}$, and $R$ due to the larger deviations therein. In summary, compared to the standard PID controller, the RL controller obtains similar precision in regulating $R$ and $Z$, but significantly extends this capability by simultaneously providing closed-loop control over minor radius $a$, upper and lower triangularity $\delta_\mathrm{u}$, $\delta_\mathrm{l}$, and elongation $\kappa$.

\section*{Data availability}
Data sets generated during the current study are available from the corresponding author on reasonable request.
\section*{Code availability}
Codes used during the current study are available from the corresponding author on reasonable request.

\bibliographystyle{naturemag}
\bibliography{ref}

\section*{Acknowledgments}
The authors would like to thank the entire HL-3 team for providing experimental data and engineering support for the verification experiments in the HL-3 tokamak. This work was supported in part by the National Key Research and Development Program of China under Grant 2024YFE03020001, the Zhejiang Provincial Natural Science Foundation of China under Grant LR23F010005, the National Nature Science Foundation of China under Grant No. U21A20440 and No. 12405253.
\section*{Author contributions}
N.N.W., Z.Y., R.L., and N.W. contributed equally to this work. R.L., M.X., Z.Z., and W.Z. conceived this collaboration. N.N.W., R.L., and N.W. led the design \& implementation of the dynamics models and reinforcement learning approach; Z.Y. and G.Z. designed and implemented EFITNN; Y.C., X.G., F.G., and B.L. contributed to the real-time on-device implementation.
N.N.W, Z.Y., R.L., and N.W. discussed the results; Q.D. and J.L. contributed to the analyses and illustration of some results. N.N.W. and R.L. wrote this manuscript; Z.Y. and N.W. revised the manuscript; all other authors reviewed the manuscript.

\section*{Competing interests}
The authors declare no competing interests.

\end{document}


\renewcommand\footnotemark{}
\renewcommand\footnoterule{}
\maketitle

\clearpage
\begin{figure*}[!t]
    \centering
    \begin{minipage}{1.0\textwidth}
        \begin{overpic}[width=1\textwidth]{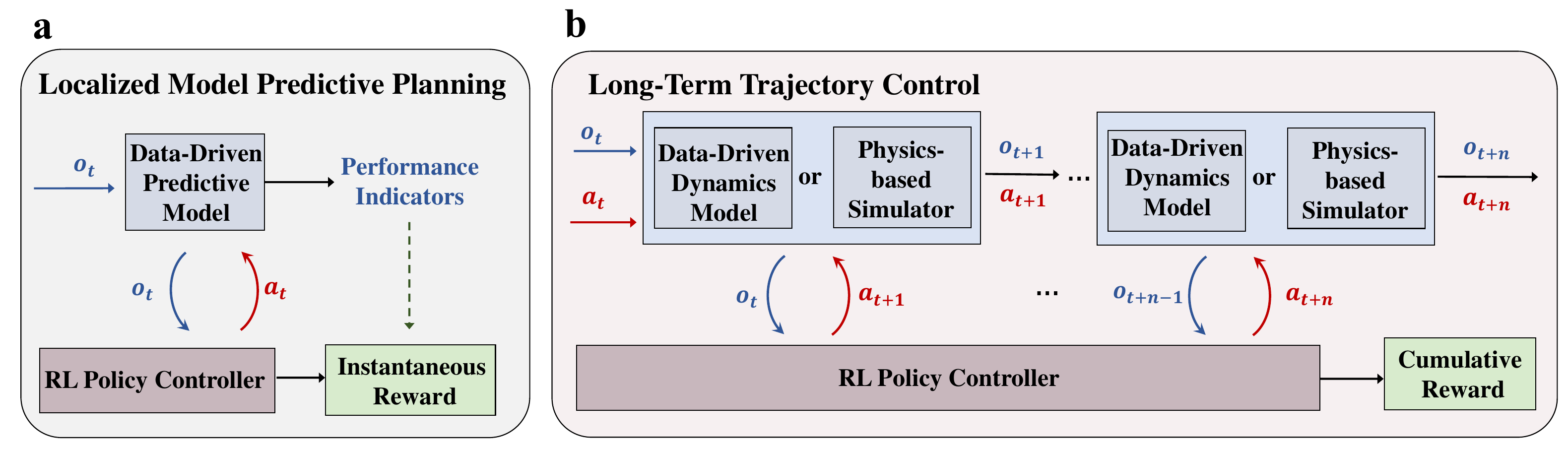}
        \end{overpic}
    \end{minipage}

    \caption{\textbf{Schematics of short-term versus long-term control planning. }\textbf{a}, Schematic diagram of the localized model-predictive planning, wherein the dynamics model refrains from autoregressive extension and guides responsive actions only. \textbf{b}, Schematic diagram of long-term trajectory control. Different from using the physics-based simulator in \cite{degrave2022magnetic}, our work utilizes the dynamics model autoregressively to facilitate training RL-based controller. }
    \label{fig:diagram_compare}
\end{figure*}

\clearpage
\begin{figure*}[!t]
    \centering
    \begin{minipage}{.95\textwidth}
        \begin{overpic}[width=1\textwidth]{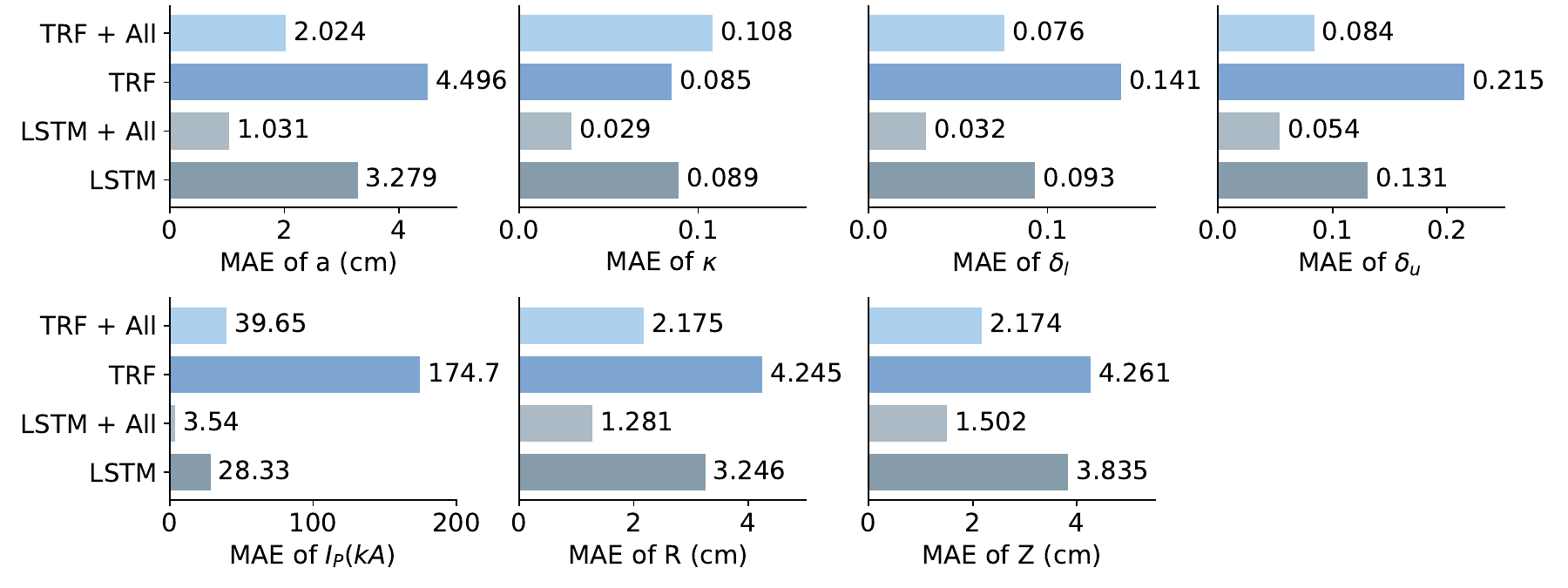}
        \end{overpic}
    \end{minipage}

    \caption{\textbf{The testing results of response model with different structures on the same test dataset}. TRF: Transformer; All: Scheduled Sampling + Ensemble + Self-Attention + Adaptive Weight.}
    \label{fig:comp_transformer}
\end{figure*}

\clearpage
\begin{figure}[!tb]
    \centering
    \includegraphics[width=.7\textwidth]{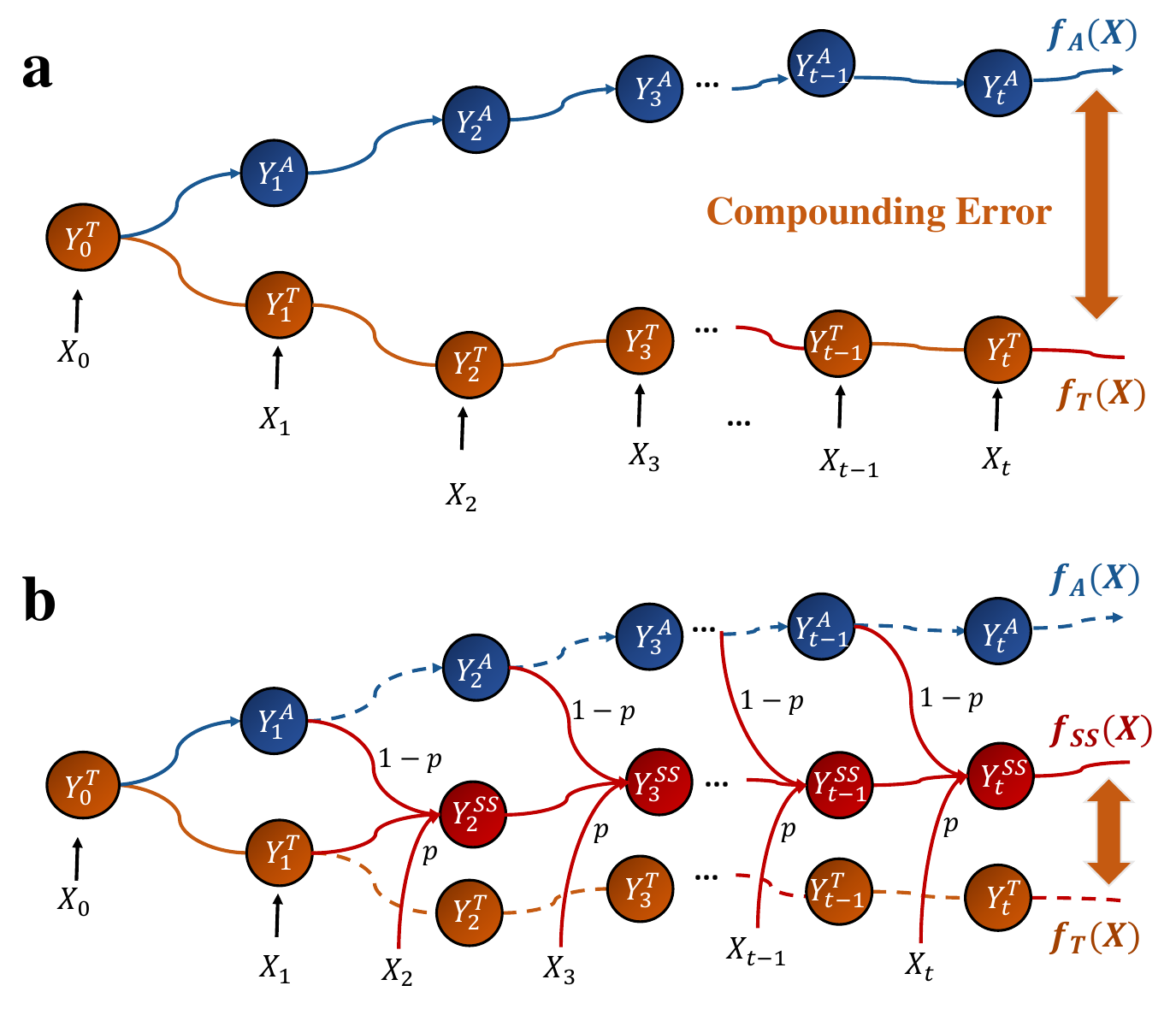}
    
\caption{\textbf{Schematic illustration of scheduled sampling.} \textbf{a}, Without any preventive training measures, the autoregressive nature of the dynamics models implies a gradually larger deviation between the autoregressive output (blue line) and the true data (orange line), due to the compounding error. \textbf{b}, The scheduling sampling technique exposes the model to real data beyond the model output, with a probabilistic teacher-forcing ratio $p$, during the training.
}
\label{fig:scheduled sampling structure}
\end{figure}

\clearpage
\begin{figure*}[!t]
    \centering

    \begin{minipage}{.95\textwidth}
        \begin{overpic}[width=1\textwidth]{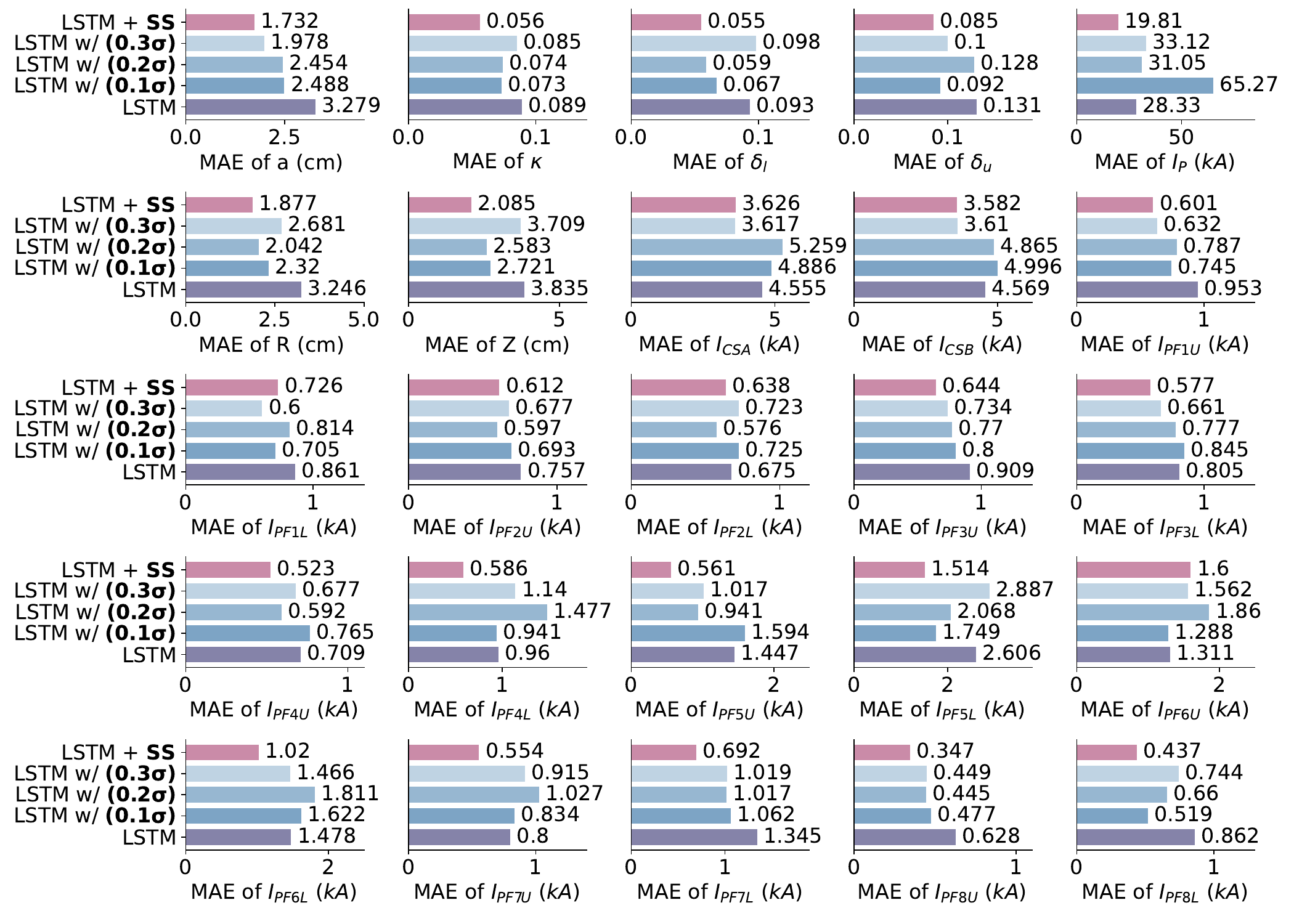}
        \end{overpic}
    \end{minipage}

    \caption{\textbf{The advantage of scheduled sampling over a noise layer}.  SS: Scheduled Sampling; LSTM w/ $(0.1\sigma$/$0.2\sigma$/$0.3\sigma)$ represents the combination of LSTM and a zero-mean Gaussian noise layer with a standard deviation of $0.1\sigma$/$0.2\sigma$/$0.3\sigma$.}
    \label{fig:trick_trans_noise}
\end{figure*}

\clearpage
\begin{figure}[!t]
    \centering
    \includegraphics[width=\textwidth]{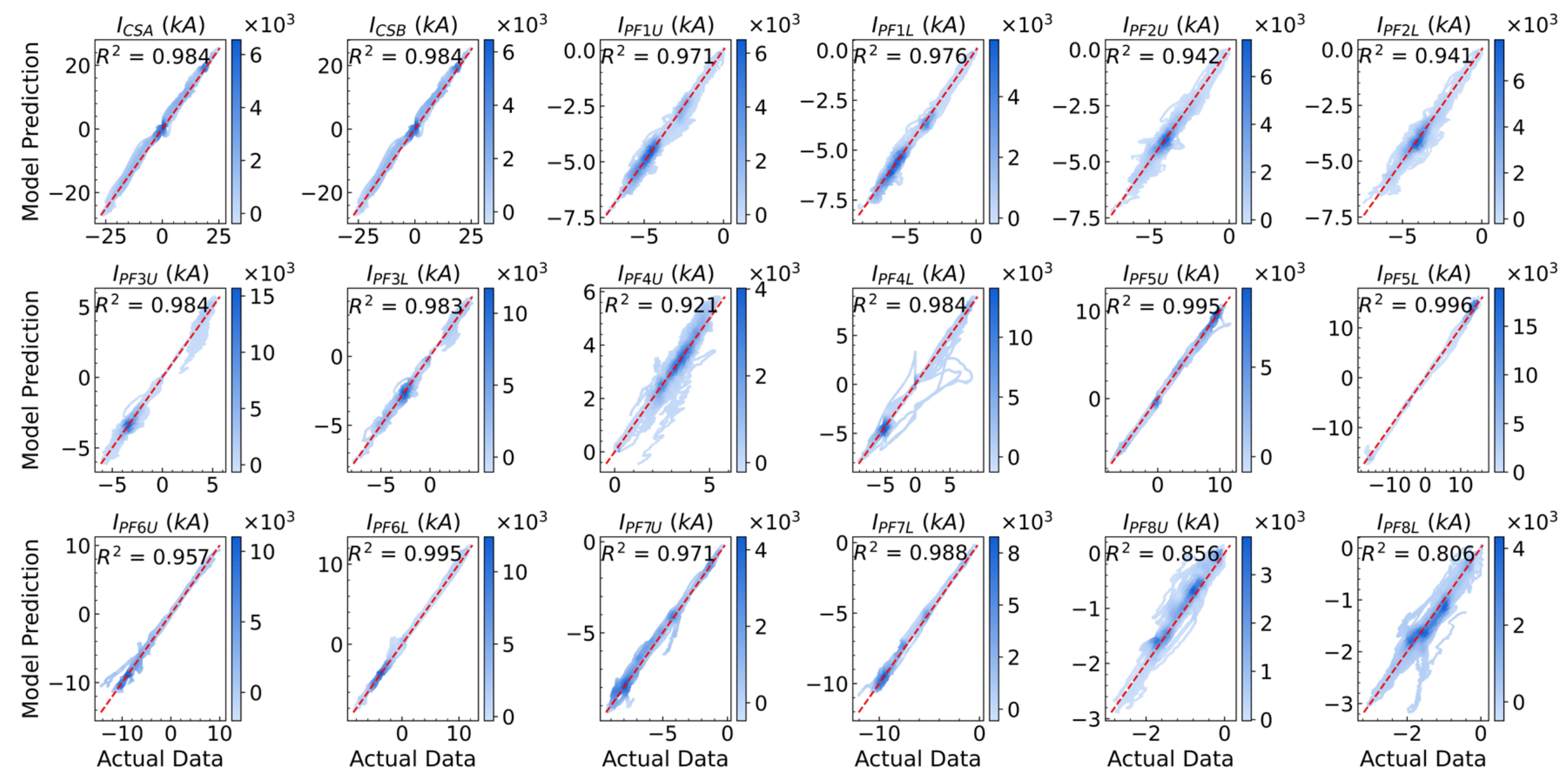}
    \caption{\textbf{Goodness-of-fit of the dynamics model for coil currents in terms of the coefficient of determination $R^2$.} The color map indicates the density of data points.
    }
    \label{fig:r2 for ic}
\end{figure}

\clearpage
\begin{figure*}[!t]
\centering
    \includegraphics[width=1.0\textwidth]{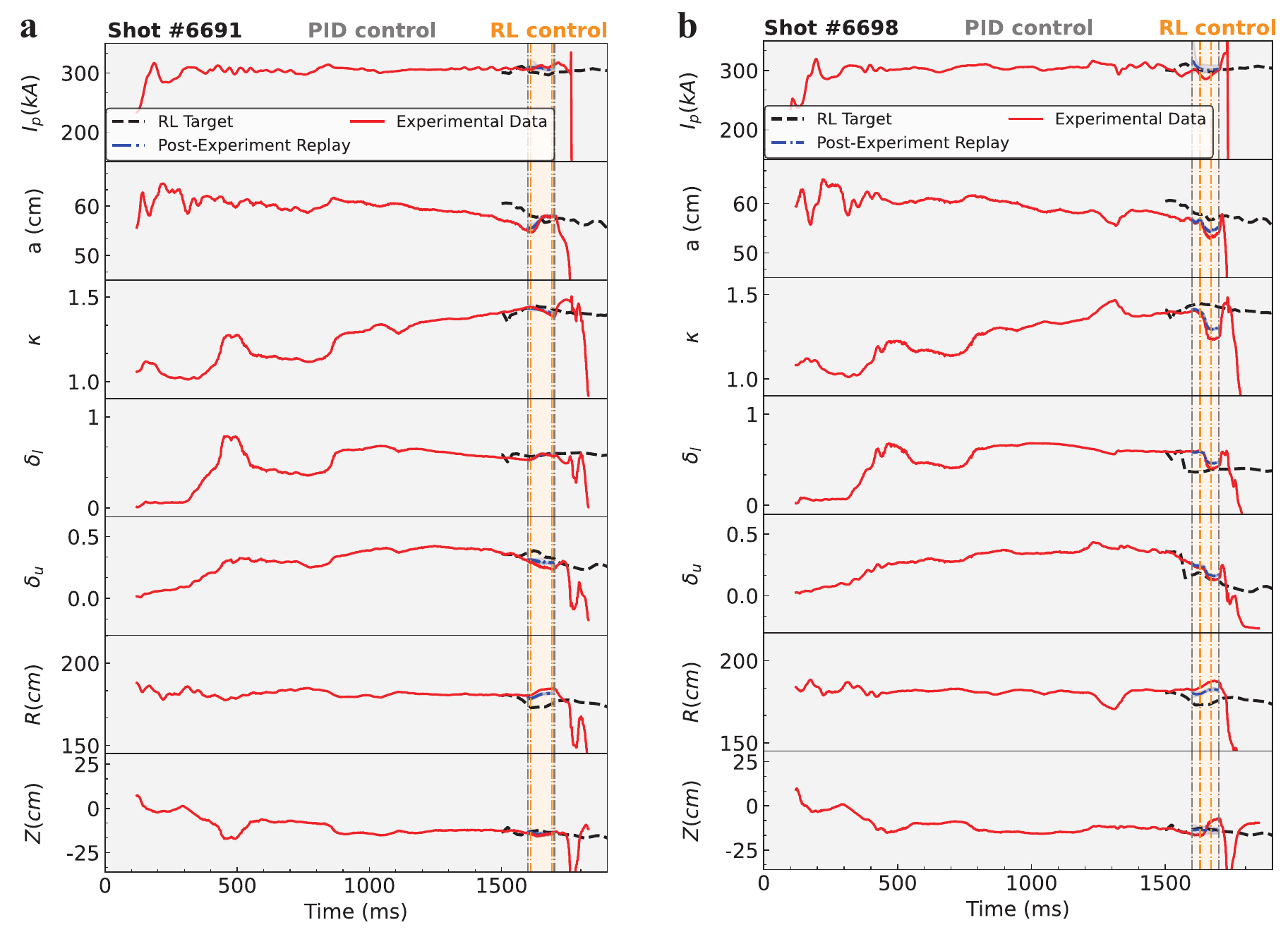}

    \caption{\textbf{Complete discharge of 7 control variables for shots \#6691 and \#6698.} In both panels, the experimental data (red solid line) is shown alongside the Reinforcement Learning (RL) target (black dashed line) and the post-experiment replay from the dynamics model (blue dashdot line). \textbf{a}, Discharge of shot \#6691, with RL control engaged from $1600$ ms to $1700$ ms, including $10$-ms transition periods at the start and end of the RL control phase. \textbf{b}, Discharge of shot \#6698, demonstrating zero-shot adaptation for changed upper and lower triangularity target, where the target values are reduced by $0.2$ compared to the training targets.
    } 
    \label{fig:full state of 6691 and 6698}
\end{figure*}

\clearpage
\begin{figure}[!tb]
    \centering
    \includegraphics[width=.5\textwidth]{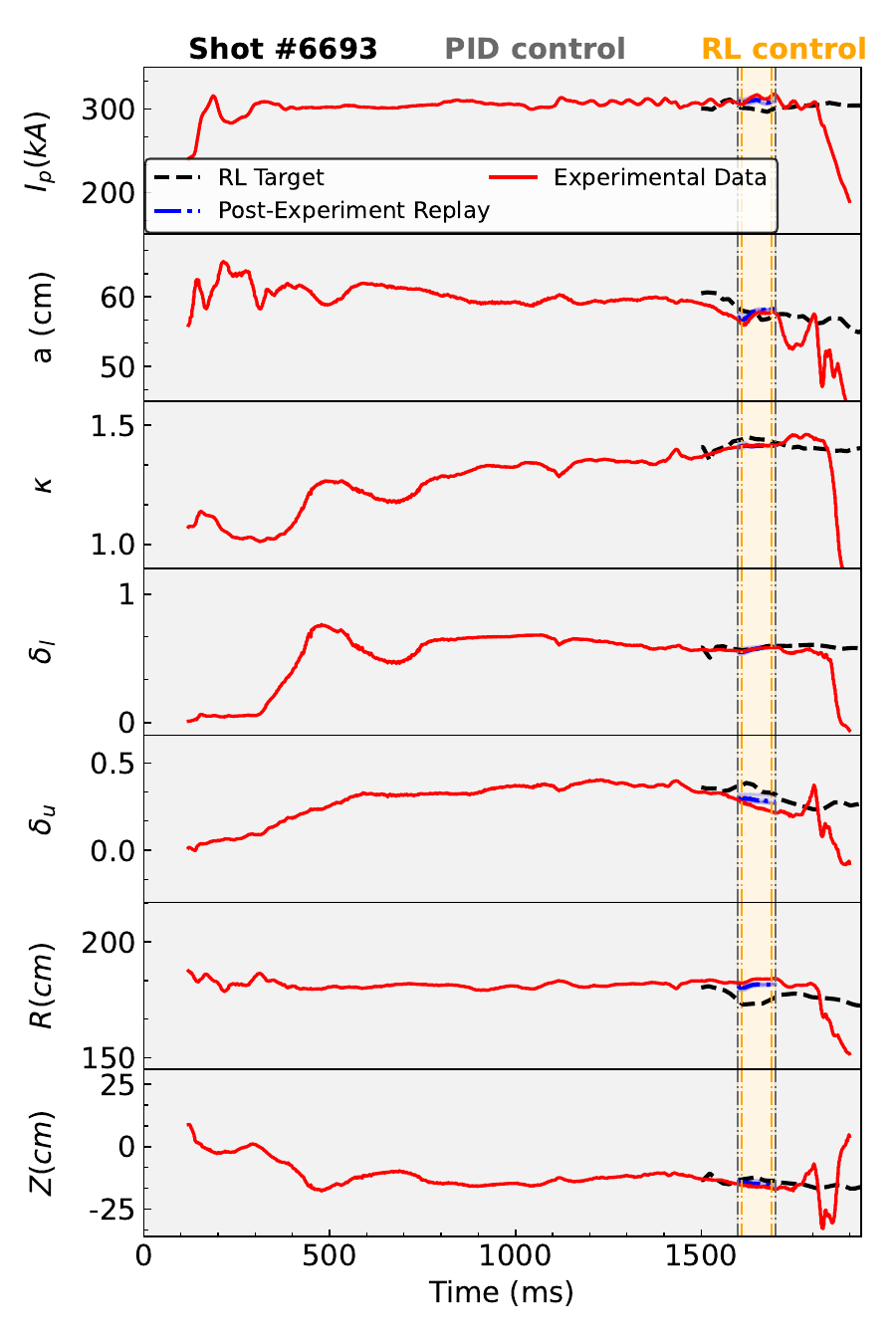}
    \caption{\textbf{Full discharge of the seven control variables for shot \#6693.} The experimental data (red solid line) is shown alongside the Reinforcement Learning (RL) target (black dashed line) and the post-experiment replay from the dynamics model (blue dashdot line). RL control is applied from $1600$ ms to $1700$ ms, including $10$ ms transition periods at the start and end.    
    }
    \label{fig:6693 state}
\end{figure}

\clearpage
\begin{figure*}[!t]
\centering
    \includegraphics[width=1.0\textwidth]{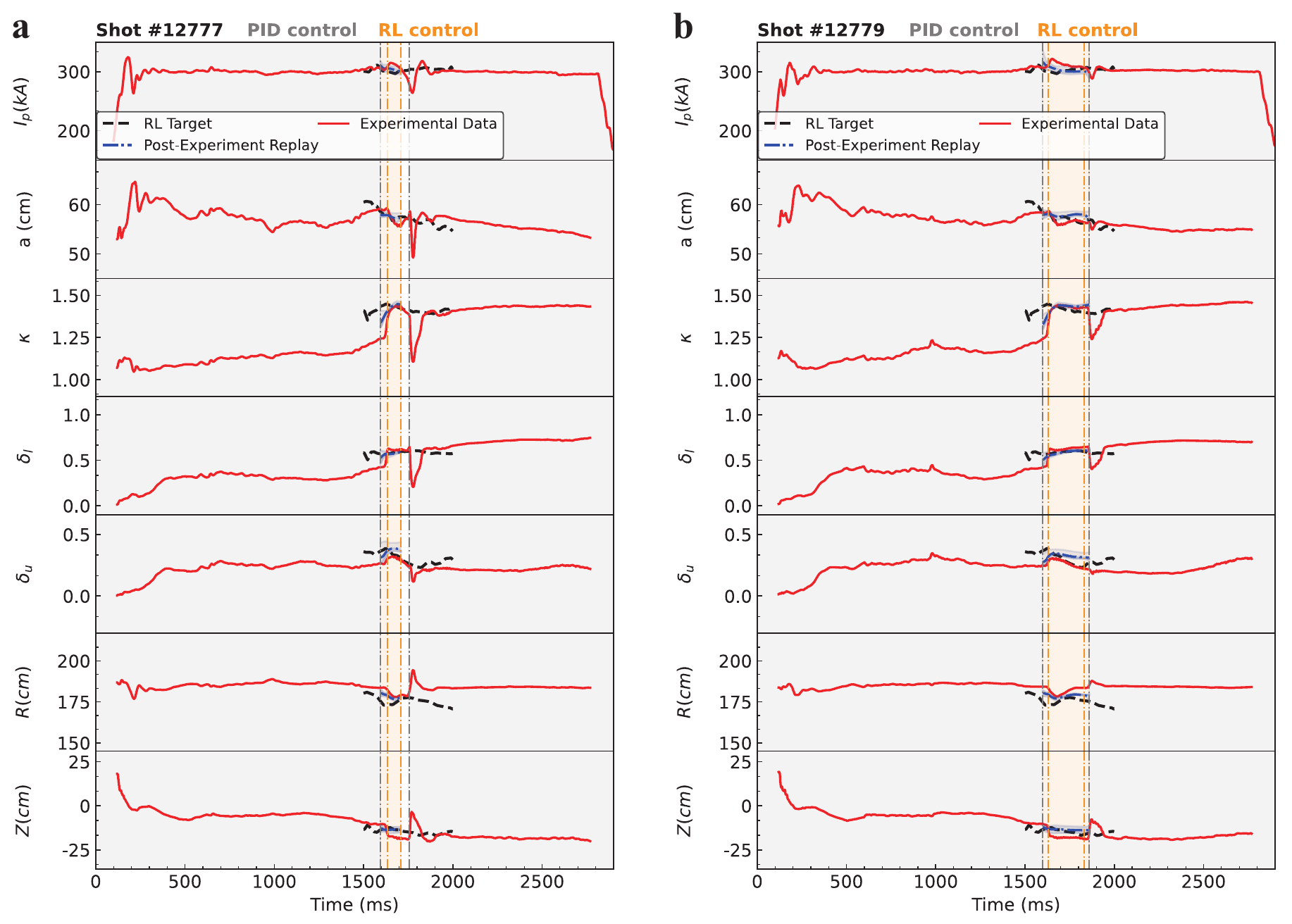}

    \caption{\textbf{Complete discharge of seven control variables for shots \#12777 and \#12779.} In both panels, the experimental data (red solid line) is shown alongside the Reinforcement Learning (RL) target (black dashed line) and the post-experiment replay from the dynamics model (blue dashdot line). \textbf{a}, Shot \#12777: RL control is engaged from $1600$ ms to $1700$ ms, with a $30$ ms transition period at the start only. From $1700$ ms to $1730$ ms, the control voltages remain constant at their $1700$ ms values. \textbf{b}, Shot \#12779: RL control is engaged from $1600$ ms to $1860$ ms, with $30$ ms transition periods at the start and end of the RL control.
    } 
    \label{fig:full state of 12777 and 12779}
\end{figure*}

\clearpage
\begin{figure}[!tb]
    \centering
    \includegraphics[width=0.5\textwidth]{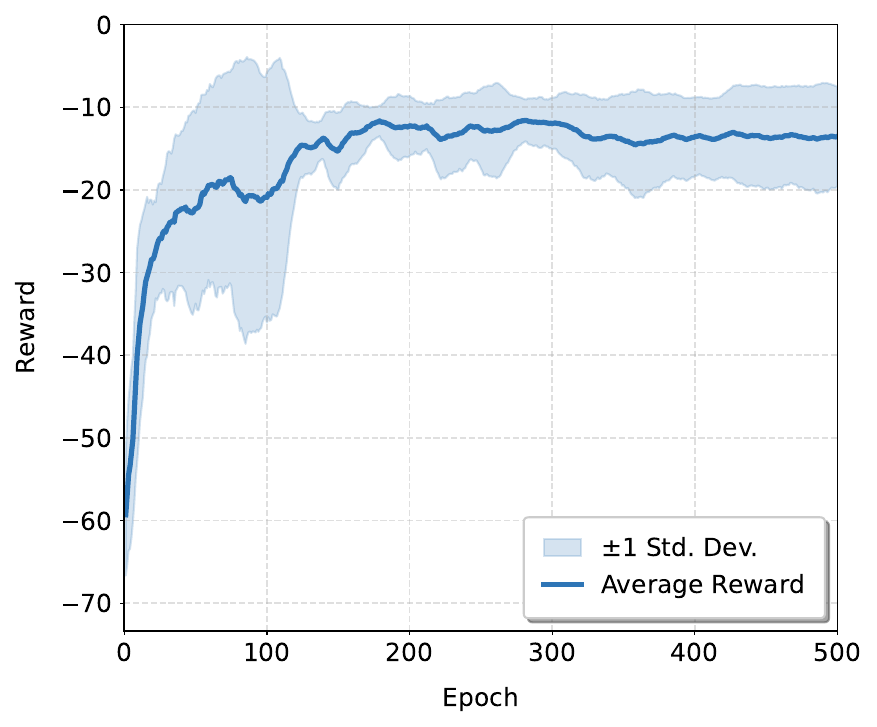}
    \caption{\textbf{Average training curve of the RL reward.} The results are averaged over $10$ training runs with different random seeds, and the shaded area represents the standard deviation across these $10$ runs. The model converges after $500$ epochs ($25,000$ interaction steps). 
    }
    \label{fig:avg_reward_rl_train}
\end{figure}

\clearpage
\begin{figure}[!tb]
    \centering
    \includegraphics[width=1\textwidth]{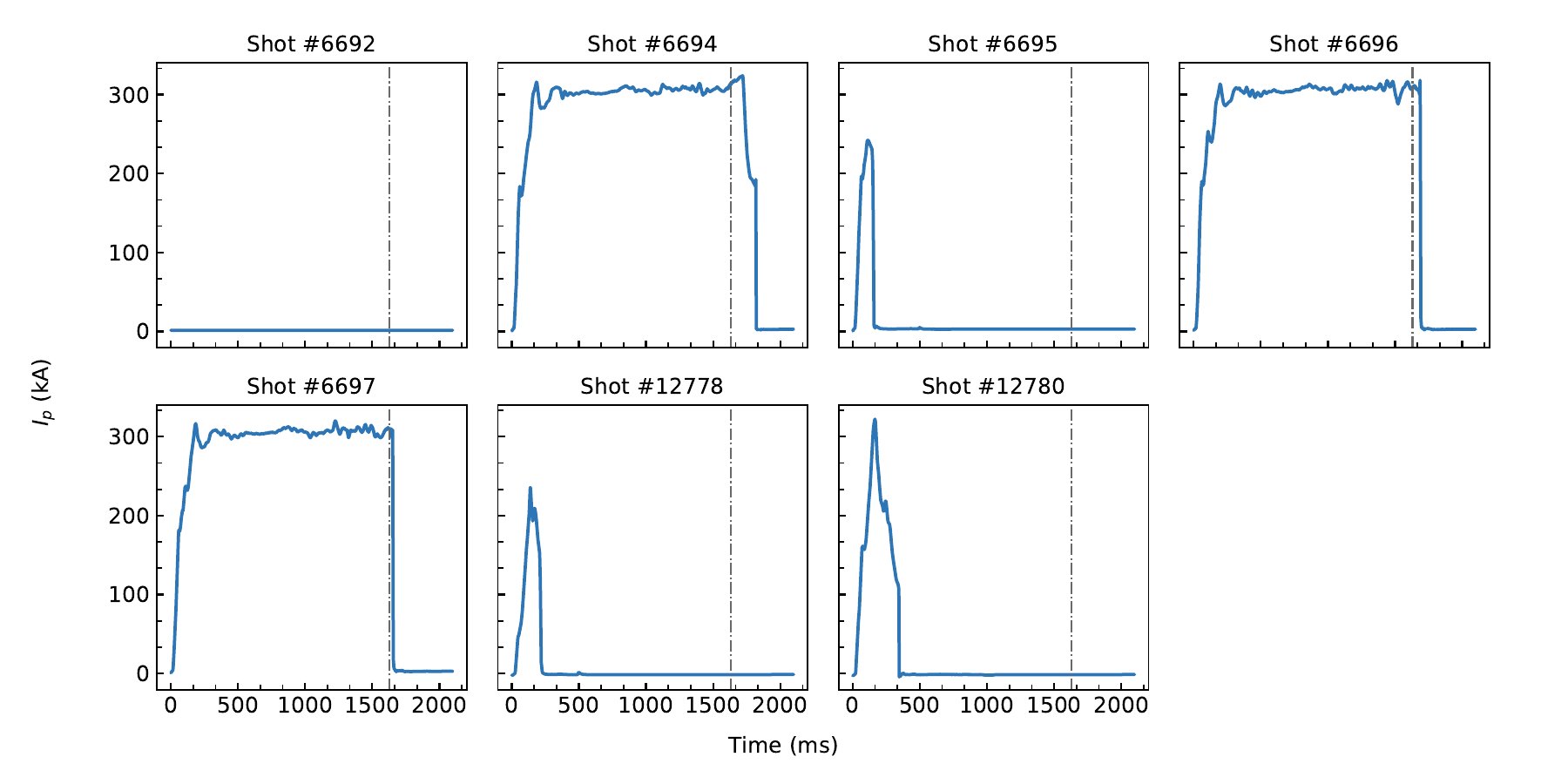}
    \caption{\textbf{Plasma current evolution for HL-3 shots \#6692–\#6697 and \#12778–\#12780.} Shots \#6694, \#6696 and \#6697 initiate ramp-down before $1600$ ms, while the other shots fail to achieve plasma initiation. 
    }
    \label{fig:6691-6698 IP}
\end{figure}

\clearpage
\begin{table*}[!t]
\centering
\footnotesize
\caption{\centering Flat-top plasma current distribution in the dataset.}
\label{data cate}
\begin{tabular}{@{}>{\centering\arraybackslash}m{3.7cm}>{\centering\arraybackslash}m{1.3cm}>{\centering\arraybackslash}m{1.3cm}>{\centering\arraybackslash}m{1.3cm}>
{\centering\arraybackslash}m{1.3cm}>{\centering\arraybackslash}m{1.3cm}>{\centering\arraybackslash}m{1.3cm}>{\centering\arraybackslash}m{1.3cm}}
\toprule
Plasma Current (kA)&$300$&$400$ &$500$&$600$&$700$&$800$&$1000$\\
\midrule
Number of Shot&$258$&$94$&$245$&$77$&$97$&$32$&$29$\\
\bottomrule
\end{tabular}
\end{table*}

\clearpage
\begin{table*}[!t]
\centering
\footnotesize
\caption{\centering Input and output variables used in dynamics model, with the length of the time series $n = 30$ and the temporal resolution $\Delta{t} = 1$ ms. As the CS coil is composed of two parallel connected sets of coils, the dimension of coil current ($I_c$) equals that of coil voltage ($U$) plus $1$.}
\label{dynamics model in and out}

\clearpage
\begin{tabular}{@{}>{\centering\arraybackslash}m{2cm}>{\centering\arraybackslash}m{2cm}>{\centering\arraybackslash}m{2cm}>{\centering\arraybackslash}m{5cm}>{\centering\arraybackslash}m{3cm}}
\toprule
Input variable&Unit&Number of channels&Description&Time\\
\midrule
$I_{p}$&kA&1&Plasma current&($t-n\Delta{t}$)$\sim$($t-\Delta{t}$)\\

$a$&cm&1&Minor radius&($t-n\Delta{t}$)$\sim$($t-\Delta{t}$)\\
$\kappa$&dimensionless&1&Elongation&($t-n\Delta{t}$)$\sim$($t-\Delta{t}$)\\
$\delta_{u}$&dimensionless&1&Upper triangularity&($t-n\Delta{t}$)$\sim$($t-\Delta{t}$)\\
$\delta_{l}$&dimensionless&1&Lower triangularity&($t-n\Delta{t}$)$\sim$($t-\Delta{t}$)\\
$R$&cm&1&Radial position of plasma geometric center&($t-n\Delta{t}$)$\sim$($t-\Delta{t}$)\\
$Z$&cm&1&Vertical position of plasma geometric center&($t-n\Delta{t}$)$\sim$($t-\Delta{t}$)\\
$I_c$&A&18&Coil current&($t-n\Delta{t}$)$\sim$($t-\Delta{t}$)\\
$U$&V&17&Coil voltage&($t-(n-1)\Delta{t}$)$\sim$($t$)\\
$TF$&T&1&Toroidal magnetic field&($t-n\Delta{t}$)$\sim$($t-\Delta{t}$)\\
$Vloop$&V&1&Loop voltage&($t-n\Delta{t}$)$\sim$($t-\Delta{t}$)\\
\bottomrule
Output variable&Unit&Number of channels&Description&Time\\
\midrule
$I_{p}$&\multicolumn{1}{c}{kA}&\multicolumn{1}{c}{1}&\multicolumn{1}{c}{Plasma current}&\multicolumn{1}{c}{$t$}\\
$a$&cm&1&Minor radius&{$t$}\\
$\kappa$&dimensionless&1&Elongation&{$t$}\\
$\delta_{u}$&dimensionless&1&Upper triangularity&{$t$}\\
$\delta_{l}$&dimensionless&1&Lower triangularity&{$t$}\\
$R$&cm&1&Radial position of plasma geometric center&{$t$}\\
$Z$&cm&1&Vertical position of plasma geometric center&{$t$}\\
$I_c$&A&18&Coil current&{$t$}\\
\bottomrule
\end{tabular}
\end{table*}

\clearpage
\begin{table}[!t]
\centering
\footnotesize
\caption{State and action space for RL-based magnetic control.}
\label{state and action space}
\centering{}
\begin{tabular}{@{}C{2.8cm}C{12.5cm}}
\toprule
Name &Description\\
\midrule
\multirow{2}{*}{State Space $\mathcal{S}$}&  Target and currently observed plasma current, and six shape and position parameters\\
& ($14$ dimensions)\\
\midrule

\multirow{2}{*}{Action Space $\mathcal{A}$}
& Voltage commands of CS and $8$ pairs of PF coils \\
& ($17$ dimension)\\
\toprule
\end{tabular}
\end{table}

\clearpage
\begin{table*}[!tb]
\centering
\footnotesize
\caption{\centering Time to convergence for model training on different devices.}
\label{RL traing time}
\begin{tabular}{@{}>{\centering\arraybackslash}m{4.6cm}>{\centering\arraybackslash}m{3.3cm}>{\centering\arraybackslash}m{3.3cm}>{\centering\arraybackslash}m{3.3cm}}
\toprule
Device &RTX 2080Ti &RTX A6000 &RTX 4090\\
\midrule
Training Time (min) &38 &29 &22\\
\bottomrule
\end{tabular}
\end{table*}

\clearpage
\bibliographystyle{naturemag}
\bibliography{supplement}